\documentclass[journal]{IEEEtran}
\usepackage{mathrsfs}
\usepackage{bbm}
\usepackage{amssymb}
\usepackage{bbding}
\usepackage{threeparttable}

\usepackage[mathcal]{euscript}

\usepackage{psfrag,calc,url,bm}

\usepackage{cite}

\usepackage{graphicx}

\usepackage{psfrag}

\usepackage{subfigure}

\usepackage{url}

\usepackage{stfloats}

\usepackage{amsmath}

\usepackage{float}

\usepackage{algorithmic}

\usepackage[ruled,linesnumbered,vlined]{algorithm2e}

\usepackage{color}

\usepackage{boxedminipage}

\usepackage{amsthm}

\usepackage{multirow}

\usepackage{setspace}

\usepackage{soul}
\usepackage{epstopdf}

\allowdisplaybreaks[3]

\usepackage{multirow}
\usepackage{amsmath}
\usepackage{booktabs}

\usepackage{makecell}
\usepackage{diagbox}
\usepackage{colortbl}
\usepackage{xcolor}
\newtheorem{rem}{Remark}

\begin{document}

\title{Spectral- and Energy-efficient Multi-BS Multi-RIS Pinching-antenna Systems: A GNN-based Approach}
\author{Changpeng He, Yang Lu,~\IEEEmembership{Senior Member,~IEEE}, Wei Chen,~\IEEEmembership{Senior Member,~IEEE}, Bo Ai,~\IEEEmembership{Fellow,~IEEE}, \\Arumugam Nallanathan,~\IEEEmembership{Fellow,~IEEE} and
Zhiguo Ding,~\IEEEmembership{Fellow,~IEEE}
\thanks{Changpeng He and Yang Lu are with the State Key Laboratory of Advanced Rail Autonomous Operation, and also with the School of Computer Science and Technology, Beijing Jiaotong University, Beijing 100044, China (e-mail: 25110135@bjtu.edu.cn, yanglu@bjtu.edu.cn).}
\thanks{Wei Chen and Bo Ai are with the School of Electronics and Information Engineering, Beijing Jiaotong University, Beijing 100044, China (e-mail: weich@bjtu.edu.cn, boai@bjtu.edu.cn).}
\thanks{Arumugam Nallanathan is with the School of Electronic Engineering and Computer Science, Queen Mary University of London, London and also with the Department of Electronic Engineering, Kyung Hee University, Yongin-si, Gyeonggi-do 17104, South Korea (e-mail: a.nallanathan@qmul.ac.uk).}
\thanks{Zhiguo Ding is with the School of Electrical and Electronic Engineering (EEE), Nanyang Technological University, Singapore 639798 (Zhiguo.ding@ntu.edu.sg).}
}

\maketitle

\begin{abstract}
This paper investigates coordinated downlink transmission in a multi-base station (multi-BS) multi-reconfigurable intelligent surface (multi-RIS)-assisted pinching-antenna (PA) system, where each user equipment (UE) is associated with a single BS and each BS is equipped with movable PAs deployed on parallel waveguides. We formulate sum rate (SR) and energy efficiency (EE) maximization problems by jointly optimizing PA placement, RIS phase shifts, transmit beamforming, and BS-UE association under constraints of inter-PA spacing, power budget, and unit-modulus phase shift. To address the resulting highly coupled mixed-variable problem, we propose a three-stage graph neural network (GNN) that integrates heterogeneous and homogeneous graph representations and is trained end-to-end in an unsupervised manner. Extensive numerical results demonstrate that the proposed three-stage GNN consistently outperforms representative system and learning baselines, generalizes well to unseen numbers of UEs, RISs, and BSs, and maintains millisecond-level inference time. Besides, the results validate the effectiveness of the proposed design from both system and architectural perspectives. Moreover, 
PAs are shown to enhance SR and EE, and the performance gain is enlarged with increasing number of PAs.
\end{abstract}


\section{Introduction}

The sixth generation (6G) wireless network is expected to support immersive services, native intelligence, and stringent spectral- and energy-efficiency requirements in highly dynamic environments, making the conventional paradigm of optimizing only transceivers increasingly inadequate \cite{wang2023road}. To enlarge the controllable degrees of freedom (DoFs) of wireless propagation, reconfigurable intelligent surfaces (RISs) have attracted extensive attention as a low-power means of shaping channels via programmable reflections \cite{di2020smart1,basar2019wireless,wu2021intelligent,liu2021reconfigurable}. RIS-aided transmission can improve coverage, spectral efficiency, energy efficiency (EE) , and physical-layer security \cite{huang2019reconfigurable,guan2020intelligent,yu2020robust,pan2021intelligent}. Nevertheless, RISs mainly provide environment-side control, and their performance is still constrained by cascaded path loss and the fixed geometry of active arrays, especially in blocked or cell-edge scenarios.

Recently, flexible-antenna technologies such as movable antennas (MAs), fluid antenna systems (FASs), and pinching-antenna (PA) systems (also known as PASS) have emerged as a complementary paradigm for wireless channel reconfiguration \cite{zhu2024movable,new2025fluid,liu2025passtutorial,ding2025pinching}. By enabling adaptive repositioning of radiating elements, these architectures exploit the spatial variation of wireless channels within a prescribed aperture and create additional DoFs beyond conventional digital beamforming. For example, MA/FAS studies have established field-response based channel models, quantified the gains of antenna position adaptation, and demonstrated substantial benefits in multi-user and multiple-input multiple-output (MIMO) systems \cite{zhu2024modeling,zhu2024multiuser,ma2024mimo,mei2024graphma}. PA systems further introduce low-loss dielectric waveguides and movable PAs, making it possible to radiate signals from different positions along extended waveguides and thus enhance line-of-sight accessibility and aperture reconfigurability \cite{liu2025passtutorial,ding2025pinching}. This motivates the integration of PA systems with RISs: PA systems offer transmitter-side spatial reconfiguration, while RISs provide environment-side wave control; together they can potentially deliver more robust coverage and higher efficiency than either mechanism alone. Their interplay is also nontrivial, as the gains brought by antenna repositioning and environmental reflection may be complementary or partially overlapping depending on the propagation geometry \cite{wei2025friendsfoes}.

The coordinated design of a multi-base station (multi-BS) multi-RIS-assisted PA system is, however, highly challenging. To avoid the stringent synchronization and signaling overhead required by coherent multi-BS transmission, each user equipment (UE) is served by a single BS, and the introduced BS-UE association serves as an additional degree of freedom for interference management and load balancing. Consequently, system performance depends on the joint optimization of PA positions, baseband beamforming, RIS phase shifts, and BS-UE association, all of which are tightly coupled through the effective channels. The resulting SR- and EE-maximization problems involve mixed continuous, unit-modulus, and binary variables under power and spacing constraints. Compared with classical beamforming design, the combination of multi-BS coordination and multi-variable optimization renders conventional iterative algorithms substantially more computationally demanding and harder to converge, with significantly degraded scalability for large antenna arrays, dense RIS deployments, or fast-varying network topologies \cite{bjornson2014optimal,pan2021intelligent,cvx-iterations}.

Learning-based wireless optimization provides a promising alternative. Prior to graph-based models, general deep architectures such as multilayer perceptrons (MLPs) and convolutional neural networks (CNNs) had already been explored for beamforming and related transmission design problems \cite{zhang2019deep,lin2019bfnn,deeptx2022,kang2024maDL}. When the problem dimension is fixed and the input admits a regular grid-like representation, these models are effective. However, they do not exploit the relational structure among network entities such as BSs and UEs, and their transfer capability is therefore often limited when the number of network entities or the interference topology changes. In contrast, graph neural networks (GNNs) are naturally suited to wireless systems because propagation, interference, and cooperation relations can be represented by graphs, enabling structure-aware policy learning with favorable scalability and permutation-equivariant inductive bias \cite{shen2020graph,shen2023gnn}. Recent studies have demonstrated the effectiveness of GNNs for beamforming and wireless network design \cite{li2024gnn,wang2024gnn}. For flexible-antenna systems, staged GNN frameworks have also started to emerge for FAS and PA systems \cite{he2025fasnet,xie2025graph,gpass2025,rispassgnn2025}. Nevertheless, existing learning-based PA systems/RISs work mainly focus on the single-cell scenarios, and omit BS-UE association. A unified framework for coordinated multi-BS multi-RIS-assisted PA systems is still lacking.

To fill this gap, this paper studies coordinated downlink transmission in a multi-BS multi-RIS-assisted PA system and develops a three-stage GNN for end-to-end unsupervised optimization. The main contributions are summarized as follows.
\begin{itemize}
    \item We formulate a multi-BS multi-RIS-assisted PA systems serving multiple UEs, based on which, we formulate SR and EE maximization problems by jointly optimizing PA placement, RIS phase design, transmit beamforming, and BS-UE association under practical power, spacing, and unit-modulus constraints. The SR and EE maximization problems are solved by a deep learning model in a unified manner.
    \item We propose a three-stage GNN architecture composed of ChanGNN, BeamGNN, and AssocGNN. By combining heterogeneous and homogeneous graph representations, the proposed model explicitly captures BS-UE, BS-RIS, RIS-UE, intra-BS, and intra-UE interactions. Moreover, feasibility-preserving readout mechanisms are designed for all decision variables, including spacing-based  placement, unit-modulus RIS phase normalization, per-BS power normalization, and differentiable BS-UE association via Gumbel-Softmax.
    \item Extensive numerical results demonstrate that the proposed framework consistently outperforms both system and model baselines. The results further show that the proposed method generalizes well to unseen numbers of UEs, BSs, and RISs while maintaining millisecond-level inference times. In addition, the ablation study confirms the importance of message passing, residual fusion, and complex-valued mappings in the proposed architecture, and the impact of key system parameters on system performance is also illustrated.
\end{itemize}

\emph{Notation}: The following mathematical notations and symbols are used throughout this paper. $\bf a$ and $\bf A$ stand for a column vector and a matrix (or tensor), respectively. The sets of real numbers, and $n$-by-$m$ real matrices are denoted by ${\mathbb{R}}$, and ${\mathbb{R}^{n \times m}}$, respectively. The sets of $n$-dimensional complex column vectors and $n$-by-$m$ complex matrices are denoted by ${\mathbb{C}^n}$ and ${\mathbb{C}^{n \times m}}$, respectively. For a complex number $a$, $\left| a \right|$ denotes its modulus and ${{\rm Re}(a)}$ denotes its real part. For a vector $\bf a$, ${\left\| \bf a \right\|}$ is the Euclidean norm. For a matrix ${\bf A}$, ${\bf A}^H$ and $  \left \|{\bf A}\right\|$ denote its conjugate transpose and Frobenius norm, respectively. $[{\bf a}]_i$, $[{\bf A}]_{i,j}$, and $[{\bf A}]_{i,:}$ denote the $i$-th element of vector ${\bf a}$, the $i$-th row and the $j$-th column element of the matrix $\bf A$, and the $i$-th row vector of the matrix $\bf A$, respectively. $\{a_{i}\}$ denotes the set of elements  for all the admissible $i$. ${\rm diag}(\cdot)$ and ${\rm Concat}(\cdot)$ denote the diagonalization and concatenation operations, respectively. $\mathcal{CN}(\cdot)$ denotes the circularly symmetric complex Gaussian distribution.

\section{Related Work}

The literature most relevant to this work can be grouped into three categories: RISs, PA systems, and GNN-based wireless optimization.

\subsection{RIS-aided Wireless Communications}
RIS-enabled wireless transmission has been extensively investigated over the past few years. Foundational works established the signal model, reflection mechanism, and fundamental advantages of RIS-assisted smart radio environments \cite{basar2019wireless,di2020smart1,wu2021intelligent,liu2021reconfigurable}. Subsequent studies considered energy-efficient design, secure transmission, and multicell coordination \cite{huang2019reconfigurable,guan2020intelligent,yu2020robust,pan2021intelligent}. These works consistently showed that jointly optimizing active beamforming and passive reflection can significantly improve system performance. Beyond pure beamforming design, association-aware RIS cellular optimization has also been studied. For example, Liu \emph{et al.} jointly optimized BS-RIS-UE association together with active and passive beamforming for RIS-aided cellular networks through a successive-access and alternating-optimization framework \cite{risassoc2021}. However, the above studies assume fixed active antenna arrays and therefore only exploit environment-side reconfiguration.

\subsection{Pinching-antenna Systems}
PA has recently emerged as a representative flexible-antenna architecture based on low-loss dielectric waveguides and movable pinching elements \cite{liu2025passtutorial,ding2025pinching}. Existing studies on PA systems  have progressed from principle and modeling toward optimization and intelligent design. On the optimization side, Bereyhi \emph{et al.} studied downlink MIMO beamforming with PA systems and jointly optimized the digital precoder and activated PA positions via fractional programming \cite{bereyhi2025downlink}. Xu \emph{et al.} further investigated joint transmit and pinching beamforming in multi-user PA systems and proposed both an optimization-based majorization-minimization and penalty dual decomposition (MM-PDD) solver and a learning-based Karush-Kuhn-Tucker-guided dual learning (KDL)-Transformer framework \cite{xu2025jointpass}. On the learning side, Xie \emph{et al.} formulated PA systems as a bipartite graph and used a graph attention network (GAT)-based model to jointly optimize antenna placement and power allocation for EE maximization \cite{xie2025graph}. Guo \emph{et al.} proposed GPASS, where one sub-GNN learns PA positions and another sub-GNN learns transmit beamforming, thus yielding a staged GNN framework tailored to pinching beamforming \cite{gpass2025}. More recently, the RIS-assisted downlink PA systems in \cite{rispassgnn2025} incorporated RIS phase design into a three-stage GNN for PA positioning, RIS configuration, and beamforming. Despite these advances, current  learning frameworks for PA systems are still mainly centered on single-BS settings and do not jointly handle coordinated multi-BS transmission and BS-UE association.

\subsection{GNN-based Wireless Optimization}
GNN-based wireless optimization has developed from general graph learning principles to increasingly coupled and structured transmission design. Early studies established why GNNs are well matched to wireless resource allocation, by showing that many wireless optimization problems exhibit graph structure and permutation equivariance \cite{shen2020graph,shen2023gnn}. Representative applications include multi-user multiple-input single-output (MU-MISO) beamforming, where Li \emph{et al.} directly mapped channel state information (CSI) graphs to beamforming vectors for SR maximization \cite{li2024gnn}, and unmanned aerial vehicle (UAV) communications, where Wang \emph{et al.} used a two-stage GNN to jointly handle UAV placement and transmission design \cite{wang2024gnn}. Multi-stage GNN design has also appeared in flexible-antenna systems; for example, He \emph{et al.} proposed a two-stage GNN for FAS, where antenna-position inference and beamforming inference are learned sequentially \cite{he2025fasnet}. More recent works have started to address coupled multi-variable optimization in RIS systems. Le \emph{et al.} designed a heterogeneous GNN for joint active and passive beamforming in distributed STAR-RIS-assisted MU-MISO systems \cite{le2024starrisgnn}, while Liu \emph{et al.} proposed a heterogeneous GNN with RIS association and beamforming updates for multi-RIS multi-user mmWave systems \cite{liu2025multirisassoc}. These studies strongly motivate graph-based learning for structured wireless optimization. Nevertheless, existing GNN frameworks have not yet unified transmitter-side spatial reconfiguration, environment-side wave control, active beamforming, and BS-UE association within one coordinated PA-RIS architecture, which is precisely the gap addressed in this paper.

\section{System Model and Problem Definition}

\begin{figure}[t]
{\centering
{\includegraphics[width=.48\textwidth]{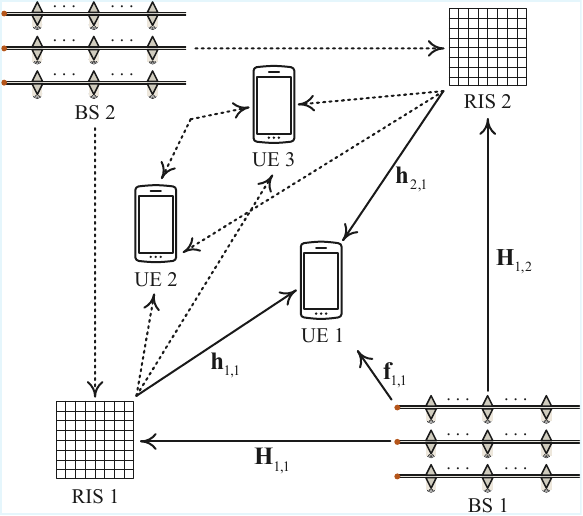}}}
\caption{Illustration of multi-BS multi-RIS-assisted PA systems for downlink multi-user communications.} 
\label{sys}
\end{figure}


As illustrated in Fig. \ref{sys}, we consider the coordinated downlink transmission for a multi-RIS-assisted PA system, where $B$ PA-BSs serve $K$ single-antenna UEs with the assistance of $R$ RISs within a rectangular region spanning a length of $D$ along the $x$-axis and $S$ along the $y$-axis.   Each BS is equipped with $N$ parallel waveguides, and $K\le N$ holds. The waveguides of BS $b$ are positioned at a height of $H_b$, extend along the $x$-axis, and span a length of $C_b$. Each waveguide hosts $M$ movable PAs. We denote the position of PA $m$ on waveguide $n$ of BS $b$ as $\bm{\psi}_{b,n,m}^{\rm P}=(x_{b,n,m}^{\rm P},y_{b,n}^{\rm P},H_b)$. UE $k$ is located at $\bm{\psi}_k^{\rm U}=(x_k^{\rm U},y_k^{\rm U},0)$ within the $S\times D$ rectangular region. RIS $r$, comprising $L$ reflecting elements, is located at $\bm{\psi}_r^{\rm R} = (x_r^{\rm R},y_r^{\rm R},z_r^{\rm R})$. For clarity, the sets of UEs, BSs, RISs, PAs per waveguide are denoted as $\mathcal{K}\triangleq\{1,\ldots,K\}$, $\mathcal{B}\triangleq\{1,\ldots,B\}$, $\mathcal{R}\triangleq\{1,\ldots,R\}$, and $\mathcal{M}\triangleq\{1,\ldots,M\}$ respectively.

\subsection{Channel Model}


The system comprises four types of channels: 1) in-waveguide propagation; 2) the PA-RIS link, 3) the RIS-UE link, and 4) the direct PA-UE link.  They are detailed as follows.

\subsubsection{In-Waveguide Propagation}
Denote the pinching beamforming matrix of BS $b$ by $\mathbf{G}_b(\{{x_{b,n,m}^{\rm P}}\})\in{\mathbb C}^{(MN) \times N}$, which is given by
\begin{flalign}\label{eqg1}
    \mathbf{G}_b&\left(\left\{{x_{b,n,m}^{\rm P}}\right\}\right) =\\
    &\begin{bmatrix}
        \mathbf{g}_{b,1} \left(\left\{{x_{b,1,m}^{\rm P}}\right\}\right)& \cdots & 0 \\
        \vdots & \ddots & \vdots \\
        0 & \cdots & \mathbf{g}_{b,N}\left(\left\{{x_{b,N,m}^{\rm P}}\right\}\right)
    \end{bmatrix},\nonumber
\end{flalign}
where 
\begin{flalign}\label{eqg2}
    \mathbf{g}_{b,n}\left(\left\{x^{\rm P}_{b,n,m}\right\}\right) = &\left[e^{-\left(\zeta+j\tfrac{2\pi}{\lambda_g}\right)\left\|{\bm\psi}^{\rm P}_{b,n,0}-{\bm\psi}^{\rm P}_{b,n,1}\right\|}, \ldots, \right.\\
    &\left.e^{-\left(\zeta+j\tfrac{2\pi}{\lambda_g}\right)\left\|{\bm\psi}^{\rm P}_{b,n,0}-{\bm\psi}^{\rm P}_{b,n,M}\right\|}\right]^T\in{\mathbb C}^M,\nonumber
\end{flalign}
with $\bm{\psi}^{\rm P}_{b,n,0} = (x_{b,n,0}^{\rm P}, y^{\rm P}_{b,n}, H_b)$ denoting the location of the feed point for waveguide $n$ of BS $b$, and $\lambda_{\rm g}=\lambda/n_{\rm eff}$ denoting the guided wavelength, where $\zeta$ denotes the in-waveguide attenuation coefficient, $\lambda$ denotes the free-space wavelength and $n_{\rm eff}$ denotes the effective refractive index of the dielectric waveguide. Notably, $\mathbf{g}_{b,n}(\{x^{\rm P}_{b,n,m}\})$ captures the in-waveguide attenuation.

\subsubsection{PA-RIS Link}

\begin{figure*}[t]
\begin{flalign}\label{ris channel1}
{\bf H}_{b,r,n}\left(\left\{{x_{b,n,m}^{\rm P}}\right\}\right)= \left[\sqrt{\frac{\beta_0}{{\|\bm{\psi}_r^{\rm R} - \bm{\psi}_{b,n,1}^{\rm P}\|^{\alpha}}}}{\bf l}_{b,r}\left({x_{b,n,1}^{\rm P}}\right),\ldots, \sqrt{\frac{\beta_0}{{\|\bm{\psi}_r^{\rm R} - \bm{\psi}_{b,n,M}^{\rm P}\|^{\alpha}}}}{\bf l}_{b,r}\left({x_{b,n,M}^{\rm P}}\right) \right]\in{\mathbb C}^{L\times M}
\end{flalign}
\hrule
\end{figure*}

The channel between $M$ PAs on waveguide $n$ of BS $b$ and RIS $r$ is modeled by \eqref{ris channel1}, where $\alpha$ denotes the path loss exponent, $\beta_0$ denotes the channel gain at the reference distance of $1$ m, and ${\bf l}_{b,r}({x_{b,n,m}^{\rm P}})\in{\mathbb C}^{L}$ denotes the small-scale fading component from PA $m$ on waveguide $n$ of BS $b$ to $L$ reflecting elements, i.e., 
\begin{flalign}\label{ris channel2}
    {\bf l}_{b,r}\left({x_{b,n,m}^{\rm P}}\right)=\sqrt{\frac{\kappa}{1+\kappa}}{\bf l}_{b,r}^{\rm LoS}\left({x_{b,n,m}^{\rm P}}\right)+\sqrt{\frac{1}{1+\kappa}}{\bf l}_{b,r}^{\rm NLoS},
\end{flalign}
where $\kappa$ denotes the Rician factor, and
\begin{flalign}\label{ris channel3}
    &{\bf l}_{b,r}^{\rm LoS}\left({x_{b,n,m}^{\rm P}}\right) =\\
    & \left[1,e^{-j\frac{2\pi}{\lambda}\Delta\varphi_{b,r,n,m}},\cdots,e^{-j\frac{2\pi}{\lambda}\left(L-1\right)\Delta\varphi_{b,r,n,m}} \right]^T,\nonumber
\end{flalign}
where $\varphi_{b,r,n,m}\in[0,2\pi)$ denotes the angle-of-departure (AoD) from PA $m$ on waveguide $n$ of BS $b$ to RIS $r$, $\Delta$ denotes the element separation, and ${\bf l}_{b,r}^{\rm NLoS}\sim\mathcal{CN}(0,\mathbf{I}_L)$ contains the non-line-of-sight (NLoS) coefficients. The channel matrix from all PAs of BS $b$ to RIS $r$ is expressed as 
\begin{flalign}\label{ris channel4}
&{\bf H}_{b,r}\left(\left\{{x_{b,n,m}^{\rm P}}\right\}\right) =\\
&{\rm Concat}\left(\left\{{\bf H}_{b,r,n}\left(\left\{{x_{b,n,m}^{\rm P}}\right\}\right)\right\}_{n=1}^N\right)\in{\mathbb C}^{L\times MN}.\nonumber
\end{flalign}

\subsubsection{RIS-UE Link}
The channel between RIS $r$ and UE $k$ is denoted by $\mathbf{h}_{r,k}{\in\mathbb C}^L$, given by  
\begin{flalign}
    \mathbf{h}_{r,k} = \sqrt{\frac{\beta_0}{{\|\bm{\psi}_r^{\rm R} - \bm{\psi}_{k}^{\rm U}\|^{\alpha}}}}\left( \sqrt{\frac{\kappa}{1+\kappa}}{\bf h}_{r,k}^{\rm LoS}+\sqrt{\frac{1}{1+\kappa}}{\bf h}_{r,k}^{\rm NLoS}\right),
\end{flalign} 
where ${\bf h}_{r,k}^{\rm LoS}$ and ${\bf h}_{r,k}^{\rm NLoS}$ denote the LoS and NLoS small-scale fading components with similar definitions to ${\bf l}_{b,r}^{\rm LoS}({x_{b,n,m}^{\rm P}})$ and ${\bf l}_{b,r}^{\rm NLoS}$.

\subsubsection{PA-UE Link}

The direct PA-UE link from all PAs of BS $b$ to UE $k$ is modeled by
\begin{flalign}
    &\mathbf{f}_{b,k}\left(\left\{{x_{b,n,m}^{\rm P}}\right\}\right) = \label{direct channel1}\\
    &\left[\mathbf{f}_{b,1,k}^T\left(\left\{{x_{b,1,m}^{\rm P}}\right\}\right), \ldots, \mathbf{f}_{b,N,k}^T\left(\left\{{x_{b,N,m}^{\rm P}}\right\}\right)\right]^T \in {\mathbb C}^{MN},\nonumber
\end{flalign}
where 
\begin{flalign}\label{direct channel2}
&\mathbf{f}_{b,n,k}\left(\left\{{x_{b,n,m}^{\rm P}}\right\}\right) =\\
    &\left[\frac{\sqrt{\eta} e^{-j\tfrac{2\pi}{\lambda}\|\bm{\psi}_k^{\rm U} - \bm{\psi}_{b,n,1}^{\rm P}\|}}{\|\bm{\psi}_k^{\rm U} - \bm{\psi}_{b,n,1}^{\rm P}\|}, \dots, \frac{\sqrt{\eta} e^{-j\tfrac{2\pi}{\lambda}\|\bm{\psi}_k^{\rm U} - \bm{\psi}_{b,n,M}^{\rm P}\|}}{\|\bm{\psi}_k^{\rm U} - \bm{\psi}_{b,n,M}^{\rm P}\|}\right]^T\in {\mathbb C}^{M},\nonumber
\end{flalign}
with $\eta = c^2/ (4 \pi f_c)^2$, where $c$ is the speed of light and $f_c$ is the carrier frequency.

\subsection{ BS-UE Association and Problem Formulation}



We assume that each UE is associated with exactly one BS, and define the corresponding variables to optimize, i.e., the association matrix, which is defined by
$\mathbf{U} \in \mathbb{R}^{B\times K}$ with binary elements $\{u_{b,k}\}$ given by
\begin{equation}
u_{b,k} =
\begin{cases}
1, & \text{UE}~k~\text{is associated with BS}~b,\\
0, & \text{otherwise.}
\end{cases}
\end{equation}
where $\sum_{b\in\mathcal{B}}u_{b,k} = 1$ holds. The UE receives the direct signal from its associated BS and the reflections from all RISs. 

The transmitted signal of each PA is a phase-shifted replica of the signal from the feed point of its waveguide. The signal emitted by PAs of BS $b$ intended for UE $k$ is given by
\begin{flalign}
    \mathbf{s}_{b,k} = \mathbf{G}_b\left(\left\{{x_{b,n,m}^{\rm P}}\right\}\right) \mathbf{w}_{b,k} u_{b,k}s_k\in\mathbb C^{MN},
\end{flalign}
where $s_k\in \mathbb{C}$ represents the information symbol for UE $k$ with $\mathbb{E}[|s_k|^2]=1$ and  $\mathbf{w}_{b,k} \in \mathbb{C}^{N}$ denotes the baseband beamforming vector of BS $b$ for UE $k$.

The received signal at UE $k$ is given by \eqref{r_pris}, 
\begin{figure*}[t]
\begin{flalign}\label{r_pris}
&\widetilde{y}_{k}= \sum_{b\in\mathcal{B}}\underbrace{\left( {{\bf f}_{b,k}^H\left(\left\{{x_{b,n,m}^{\rm P}}\right\}\right)+\sum_{r\in\mathcal{R}}{\bf{h}}_{r,k}^H{\bm\Phi}_r {\bf H}_{b,r}\left(\left\{{x_{b,n,m}^{\rm P}}\right\}\right)} \right)}_{\triangleq \widetilde{\bf h}_{b,k}^H\left(\left\{x^{\rm P}_{b,n,m}\right\}, \{{\bm\Phi}_r\}\right)}\sum_{{k^{\prime}}\in {\cal K}}\mathbf{G}_b\left(\left\{{x_{b,n,m}^{\rm P}}\right\}\right) \mathbf{w}_{b,k^{\prime}} u_{b,{k^{\prime}}}s_{k^{\prime}} + {n}_{k}
\end{flalign}
\hrule
\end{figure*}
where the diagonal matrix 
\begin{flalign}\label{phi_init}
{\bm\Phi}_r \triangleq {\rm diag}\left\{\left[e^{j\phi_{r,1}},e^{j\phi_{r,2}},...,e^{j\phi_{r,L}}\right]\right\}\in {\mathbb C}^{L\times L}
\end{flalign}
denotes the phase-shift matrix of RIS $r$ with $\phi_{r,l} \in [0,2\pi)$, and $n_{k}\sim\mathcal{CN}(0,{\sigma_{k}^2})$ denotes the additive white Gaussian noise (AWGN). The achievable rate at UE $k$ is given by \eqref{r_u_p}.

\begin{figure*}[t]    
\begin{flalign}\label{r_u_p}
&{R}_k\left(\left\{x^{\rm P}_{b,n,m}\right\}, {\left\{ {{{\bf{w}}_{b,k}}} \right\},\left\{{\bm\Phi}_r\right\}, \mathbf{U}} \right) =\\
&{\log _2}\left( {1 + \frac{{{{\left| {\sum_{b\in\mathcal{B}}u_{b,k}\widetilde{\bf h}_{b,k}^H\left(\left\{x^{\rm P}_{b,n,m}\right\}, \{{\bm\Phi}_r\}\right)\mathbf{G}_b\left(\left\{{x_{b,n,m}^{\rm P}}\right\}\right){{\bf{w}}_{b,k}}} \right|}^2}}}{{\sum_{b\in\mathcal{B}} \sum\nolimits_{k^{\prime} \in \mathcal{K}\backslash\{k\}} {\left|u_{b,k^{\prime}} {{\widetilde{\bf h}_{b,k}^H\left(\left\{x^{\rm P}_{b,n,m}\right\}, \{{\bm\Phi}_r\}\right)}\mathbf{G}_b\left(\left\{{x_{b,n,m}^{\rm P}}\right\}\right){{\bf{w}}_{b,k^{\prime}}}} \right|}^2 + \sigma _k^2}}} \right)\nonumber
\end{flalign}
\hrule
\end{figure*}

Our objective is to jointly optimize PA positions, baseband beamforming vectors, RIS reflecting coefficients, and BS-UE association to maximize the sum rate (SR) or EE:
\begin{subequations}\label{p1}
\begin{align}
    {\rm P1}: &\max_{\substack{
    \left\{x^{\rm P}_{b,n,m}\right\},\mathbf{U},\\\left\{{\bf w}_{b,k}\right\},
    \left\{{\bm\Phi}_r\right\}}}   
    {\sum\nolimits_{k\in\mathcal{K}} {R}_k\left(\left\{x^{\rm P}_{b,n,m}\right\}, {\left\{ {{{\bf{w}}_{b,k}}} \right\},\left\{{\bm\Phi}_r\right\}, \mathbf{U}} \right)} \label{p1:2a1}\\
    {\rm P2}: &\max_{\substack{
    \left\{x^{\rm P}_{b,n,m}\right\},\mathbf{U},\\\left\{{\bf w}_{b,k}\right\},
    \left\{{\bm\Phi}_r\right\}}} 
    \frac{\sum\limits_{k\in\mathcal{K}} {R}_k\left(\left\{x^{\rm P}_{b,n,m}\right\}, {\left\{ {{{\bf{w}}_{b,k}}} \right\},\left\{{\bm\Phi}_r\right\}, \mathbf{U}} \right)}{\sum_{b\in\mathcal{B}}\sum_{k\in\mathcal{K}} u_{b,k}\|\mathbf{w}_{b,k}\|^2 + P_{\rm C}} \label{p1:2a2}\\
    \rm{s.t.}~& 0 \le x_{b,n,m}^{\rm P} \le C_b, ~\forall b, n,m, \label{p1:2b} \\
    & x_{b,n,m}^{\rm P} - x_{b,n,m-1}^{\rm P} \ge \Delta_{\min}, ~\forall b,  n, \forall m>1, \label{p1:2c} \\
    & \sum\nolimits_{k\in\mathcal{K}} u_{b,k}\|\mathbf{w}_{b,k}\|^2 \le P_{\max}, ~\forall b, \label{p1:2d} \\
    &\phi_{r,l} \in [0,2\pi), ~\forall r, l, \label{p1:2e} \\
    &\sum\nolimits_{b\in\mathcal{B}}u_{b,k} = 1, ~\forall k, \label{p1:2f}\\
    &u_{b,k}\in\{0,1\}, ~\forall k, b, \label{p1:2g}
\end{align}
\end{subequations}
where ${P_{\rm C}}$ denotes the constant circuit power, $\Delta_{\min} > 0$ denotes the minimum spacing between any two adjacent PAs, and ${P_{\rm max}}$ denotes the total power budget.

Notably, the considered Problems ${\rm P}_1$ and ${\rm P}_2$ are challenging to solve due to complexly coupled variables and integer variables. Specifically, they deviate significantly from the tractable forms amenable to conventional convex optimization approaches. Instead, we propose a unified deep learning-based approach to obtain  near-optimal solutions to Problems ${\rm P}_1$ and ${\rm P}_2$.

\section{Three-Stage GNN for Joint Optimization}

\begin{figure*}[t]
{\centering
{\includegraphics[width=1\textwidth]{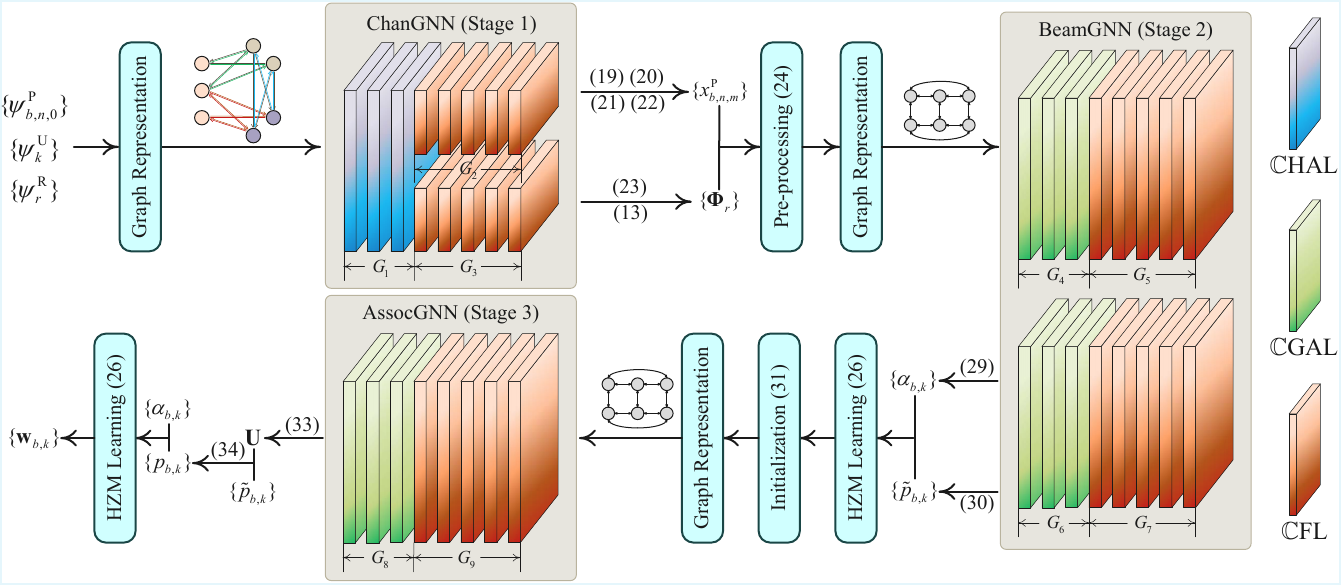}}}
\caption{Structure of the proposed three-stage GNN. \textbf{Stage 1} (ChanGNN) employs $\mathbb{C}$HAL and $\mathbb{C}$FL to jointly map location features to PA positions $\{x_{b,n,m}^{\rm P}\}$ (from BS node outputs) and RIS phase shift matrices $\{\bm{\Phi}_r\}$ (from RIS node outputs). \textbf{Stage 2} (BeamGNN) uses two parallel branches of $\mathbb{C}$GAL and $\mathbb{C}$FL to map effective channel features to hybrid coefficients $\{\alpha_{b,k}\}$ and power allocation $\{\widetilde{p}_{b,k}\}$. \textbf{Stage 3} (AssocGNN) employs $\mathbb{C}$GAL and $\mathbb{C}$FL to map beamforming gain features to the association matrix $\mathbf{U}$ via Gumbel-Softmax. Finally, $\{\mathbf{w}_{b,k}\}$ is recovered via HZM.}
\label{structure}
\end{figure*}

To exploit the spatial relationships and interference structure inherent in the considered multi-BS multi-RIS PA system, we propose a three-stage GNN that maps the locations of BSs, RISs, and UEs to PA positions, RIS phase shifts, beamforming vectors, and BS-UE association coefficients, with the objective of maximizing the system SR or EE. The proposed model employs three types of layers: the Complex Heterogeneous Graph Attention Layer ($\mathbb{C}$HAL), the Complex Graph Attention Layer ($\mathbb{C}$GAL), and the Complex Fully-Connected Layer ($\mathbb{C}$FL), which will be introduced in detail later.
 

\subsection{Graph Representation and Overall Framework}

The proposed model is a three-stage GNN, as illustrated in Figure~\ref{structure}. To facilitate effective feature extraction, we model the considered system as distinct corresponding to each of the three stages. 

\subsubsection{Stage 1} 
\begin{figure}[t]
\centering
\subfigure[Graph representation of heterogeneous graph $\mathcal{G}_1$ for Stage 1.]{\label{graphical representation1} \includegraphics[ width=0.46\linewidth]{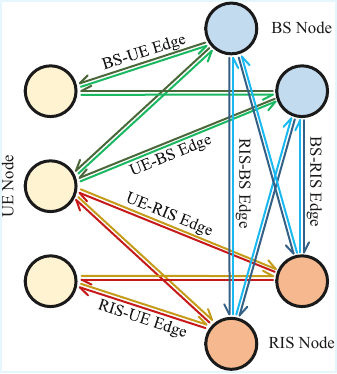}}
\subfigure[Graph representation of homogeneous graph $\mathcal{G}_2$ for Stage 2 and 3.]{\label{graphical representation2}  \includegraphics[ width=0.46\linewidth]{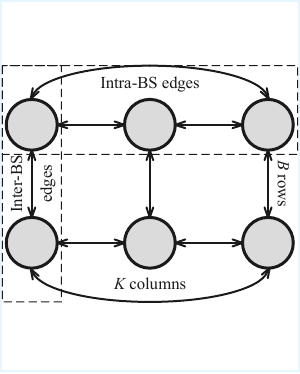}}
\caption{An example of graph representation with $K=3$, $B=2$ and $R=2$.}
\end{figure}

A heterogeneous graph $\mathcal{G}_1 = (\mathcal{V}_1, \mathcal{E}_1)$ is constructed as shown in Figure~\ref{graphical representation1}, where $\mathcal{V}_1 = \mathcal{V}_{\rm B} \cup \mathcal{V}_{\rm U} \cup \mathcal{V}_{\rm R}$ with $|\mathcal{V}_{\rm B}|=B$, $|\mathcal{V}_{\rm U}|=K$, $|\mathcal{V}_{\rm R}|=R$ denotes BS, UE, and RIS nodes, respectively. These nodes are featured by their locations, i.e., $\{\bm{\psi}_{b,n,0}^{\rm P}\}, \{\bm{\psi}_k^{\rm U}\}$ and $\{\bm{\psi}_r^{\rm R}\}$. The edge set $\mathcal{E}_1$ comprises six types of fully-connected directed edges, representing $B\times K$ BS-UE pairs, $B\times K$ UE-BS pairs, $B\times R$ BS-RIS pairs, $B\times R$ RIS-BS pairs, and $R\times K$ RIS-UE pairs, and $R\times K$ UE-RIS pairs, respectively. 

Stage 1 adopts a model termed, channel GNN  (ChanGNN), to jointly learn PA positions $\{x_{b,n,m}^{\rm P}\}$ (from BS node outputs) and RIS phase shift matrices $\{\bm{\Phi}_r\}$ (from RIS node outputs) over the defined graph.

\subsubsection{Stage 2} A  homogeneous directed graph $\mathcal{G}_2 = (\mathcal{V}_2, \mathcal{E}_2)$ is constructed as illustrated in Figure \ref{graphical representation2}, where $\mathcal{V}_2$ contains $K\times B$ nodes, each representing a BS-UE link. Each node is characterized by the effective channel (cf. \eqref{effective_channel}), computed with the obtained  $\{x_{b,n,m}^{\rm P}\}$ and $\{\bm{\Phi}_r\}$. The edge set $\mathcal{E}_2$ comprises $BK(K+B-2)/2$ bidirectional edges that characterize both inter-BS and intra-BS relationships. Specifically, the inter-BS edges for UE $k$ connect all pairs of nodes $(k,b)$ and $(k,b^{\prime})$ with $b\neq b^{\prime}$, capturing inter-BS competition for associating with the UE; while the intra-BS edges for BS $b$ connect all pairs of nodes $(k,b)$ and $(k^{\prime},b)$ with $k\neq k^{\prime}$, capturing inter-UE interference within the BS.



Stage 2 adopts a model, termed beamforming GNN (BeamGNN), to learn  $B\times K$ beamforming vectors $\{\mathbf{w}_{b,k}\}$ over the defined graph, where each node feature is mapped to a corresponding beamforming vector.

\subsubsection{Stage 3} The modeled graph is also based on $\mathcal{G}_2 = (\mathcal{V}_2, \mathcal{E}_2)$, but each node is characterized by the beamforming gain (cf. \eqref{beamgain}). 

Stage 3 adopts a model, termed association GNN (AssocGNN), to learn the BS-UE association matrix $\mathbf{U}$ from beamforming gain features.


The three GNNs are sequentially stacked and jointly trained, with the detailed processes of each stage provided as follows.


\subsection{Stage 1: ChanGNN}

The input of ChanGNN is $\{\bm{\psi}_{b,n,0}^{\rm P}\}, \{\bm{\psi}_k^{\rm U}\}$ and $\{\bm{\psi}_r^{\rm R}\}$, and it yields $\{x_{b,n,m}^{\rm P}\}$ based on the updated features of BS nodes and $\{\bm{\Phi}_r\}$ based on the updated features of RIS nodes   in a node-wise manner.

\subsubsection{Node Feature Initialization}

The initial feature vectors are defined per node type as
\begin{flalign}
 &\left[\mathbf{V}_0^{\rm B}\right]_{b,:} = \operatorname{Concat}\left(\{\bm{\psi}_{b,n,0}^{\rm P}\}_{n}\right),\\
 &\left[\mathbf{V}_0^{\rm U}\right]_{k,:} = \bm{\psi}_k^{\rm U},~\left[\mathbf{V}_0^{\rm R}\right]_{r,:} = \bm{\psi}_r^{\rm R}.\nonumber
\end{flalign}

\subsubsection{Feasibility-Guaranteed PA Placement}

To guarantee that the learned PA positions satisfy constraints \eqref{p1:2b} and \eqref{p1:2c}, the auxiliary inter-antenna spacing variables are introduced as
\begin{flalign}\label{delta}
&\delta_{b,n,1} = x_{b,n,1}^{\rm P},\nonumber\\
&\delta_{b,n,m} = x_{b,n,m}^{\rm P} - x_{b,n,m-1}^{\rm P} - \Delta_{\min},~\forall b,n,\forall m>1,
\end{flalign}
with maximum available spacing $\delta_{b,\max} = C_b - (M-1)\Delta_{\min}$. Then, constraints \eqref{p1:2b} and \eqref{p1:2c} are equivalently expressed as
\begin{flalign}\label{cons_pa}
\delta_{b,n,m} \ge 0,~ \sum\nolimits_{m^{\prime}\in\cal{M}}\delta_{b,n,m^{\prime}} \le \delta_{b,\max},~\forall b,n,m.
\end{flalign}

Notably, the reformulation allows ChanGNN to first yield $\{\delta_{b,n,m}\}$ and then reconstruct $\{x_{b,n,m}^{\rm P}\}$.

\subsubsection{$\mathbb{C}$HAL and $\mathbb{C}$FL in ChanGNN}

We employ $G_1$ $\mathbb{C}$HALs to extract heterogeneous relational features from the location inputs. Denote the output feature matrices of BS and RIS nodes  of the $G_1$-th $\mathbb{C}$HAL as $\mathbf{V}_{G_1}^{\rm BS}$ and $\mathbf{V}_{G_1}^{\rm RIS}$, respectively. Then, we employ $G_2$ $\mathbb{C}$FLs and $G_3$ $\mathbb{C}$FLs to map $\mathbf{V}_{G_1}^{\rm BS}$ and $\mathbf{V}_{G_1}^{\rm RIS}$ to $\mathbf{V}^{\prime\prime}_{G_2}$ and $\mathbf{V}^{\prime\prime}_{G_3}$, respectively, which are further used to generate $\{x_{b,n,m}^{\rm P}\}$ and $\{\bm{\Phi}_r\}$, respectively.


\paragraph{PA Position Output}
For BS $b$, we first obtained $\overline{{\delta}}_{b,n,m}$ from $\mathbf{V}^{\prime\prime}_{G_2}$ as 
\begin{flalign}
\overline{{\delta}}_{b,n,m} = \mathrm{Re}\left(\left[\mathbf{V}^{\prime\prime}_{G_2}\right]_{b,(n-1)M+m}\right).
\end{flalign}
To ensure \eqref{cons_pa}, we apply the sigmoid activation $\sigma(\cdot)$ and numerical scaling to $\overline{{\delta}}_{b,n,m}$:
\begin{flalign}
\delta_{b,n,m} = \delta_{b,\max}\sigma(\overline{\delta}_{b,n,m}),
\end{flalign}
\begin{flalign}
\delta_{b,n,m} := \left\{\begin{array}{ll}
\delta_{b,n,m}, & \displaystyle\sum_{m^{\prime}\in\cal{M}}\delta_{b,n,m^{\prime}}\le\delta_{b,\max}\\[4pt]
\dfrac{\delta_{b,n,m}}{\sum_{m^{\prime}\in\cal{M}}\delta_{b,n,m^{\prime}}}\delta_{b,\max}, & \text{otherwise}
\end{array}\right..
\end{flalign}
Then, PA positions are recovered via
\begin{flalign}\label{sum_x}
x_{b,n,m}^{\rm P} = \sum\nolimits_{m^{\prime}=1}^m\delta_{b,n,m^{\prime}}+(m-1)\Delta_{\min},~\forall b,n,m,
\end{flalign}
satisfying constraints \eqref{p1:2b} and \eqref{p1:2c}.

\paragraph{RIS Phase Shift Output}
To satisfy constraint \eqref{p1:2e}, each element of $\mathbf{V}^{\prime\prime}_{G_3}$ is normalized as
\begin{flalign}\label{cons-phi}
e^{j\phi_{r,l}} = \frac{\left[\mathbf{V}^{\prime\prime}_{G_3}\right]_{r,l}}{\left|\left[\mathbf{V}^{\prime\prime}_{G_3}\right]_{r,l}\right|},~\forall r,l,
\end{flalign}
and $\bm{\Phi}_r$ is constructed following \eqref{phi_init}.

\subsection{Stage 2: BeamGNN}

The input and output of BeamGNN are the effective channel features computed with the obtained $\{x_{b,n,m}^{\rm P}\}$ and $\{\bm{\Phi}_r\}$ and the beamforming vectors $\{\mathbf{w}_{b,k}\}$, respectively.

\subsubsection{Pre-processing}

With the obtained $\{x_{b,n,m}^{\rm P}\}$ and $\{\bm{\Phi}_r\}$, the effective channel for BS-UE node $(b,k)$ is computed as
\begin{flalign}\label{effective_channel}
&\widehat{\mathbf{h}}_{b,k}^H\left(\left\{x_{b,n,m}^{\rm P}\right\},\{\bm{\Phi}_r\}\right)=\left(\mathbf{f}_{b,k}^H\left(\left\{x_{b,n,m}^{\rm P}\right\}\right)+\right.\\
&\quad\quad\left.\sum\nolimits_{r\in\mathcal{R}}\mathbf{h}_{r,k}^H\bm{\Phi}_r\mathbf{H}_{b,r}\left(\left\{x_{b,n,m}^{\rm P}\right\}\right)\right)\mathbf{G}_b\left(\left\{x_{b,n,m}^{\rm P}\right\}\right).\nonumber
\end{flalign}
The feature of BS-UE node $(b,k)$ is initialized with the stacked effective channel:
\begin{flalign}
[\mathbf{V}^{\prime}_0]_{(b-1)K+k,:} = \widehat{\mathbf{h}}_{b,k}^H\left(\left\{x_{b,n,m}^{\rm P}\right\},\{\bm{\Phi}_r\}\right)\in\mathbb{C}^{BK\times N}.
\end{flalign}

\subsubsection{HZM Learning}

To reduce output dimensionality, we adopt the hybrid zero-forcing and maximum ratio transmission (HZM) learning \cite{hyb}, decomposing each beamforming vector as
\begin{flalign}\label{HZM}
\mathbf{w}_{b,k} = \sqrt{p_{b,k}}\,\overline{\mathbf{w}}_{b,k}(\alpha_{b,k}),~\|\overline{\mathbf{w}}_{b,k}(\alpha_{b,k})\|^2=1,
\end{flalign}
where $p_{b,k}\in\mathbb{R}^+$ is the allocated power and $\alpha_{b,k}\in[0,1]$ is the hybrid coefficient. The unit-norm direction vector is
\begin{flalign}
\overline{\mathbf{w}}_{b,k}(\alpha_{b,k}) = \frac{\alpha_{b,k}\frac{\mathbf{q}_{b,k}}{\|\mathbf{q}_{b,k}\|}+(1-\alpha_{b,k})\frac{\widehat{\mathbf{h}}_{b,k}}{\|\widehat{\mathbf{h}}_{b,k}\|}}{\left\|\alpha_{b,k}\frac{\mathbf{q}_{b,k}}{\|\mathbf{q}_{b,k}\|}+(1-\alpha_{b,k})\frac{\widehat{\mathbf{h}}_{b,k}}{\|\widehat{\mathbf{h}}_{b,k}\|}\right\|},
\end{flalign}
where $\mathbf{q}_{b,k}$ is the $k$-th column of $\mathbf{Q}_b \triangleq \mathbf{Z}_b^H(\mathbf{Z}_b\mathbf{Z}_b^H)^{-1}$ with
\begin{flalign}
\mathbf{Z}_b \triangleq \left[\widehat{\mathbf{h}}_{b,1}^H;\widehat{\mathbf{h}}_{b,2}^H;\ldots;\widehat{\mathbf{h}}_{b,K}^H\right]\in\mathbb{C}^{K\times N}.
\end{flalign}

Notably, the HZM learning reduces the required output dimension from $KBN$ complex scalars to $2KB$ real scalars, enhancing both expressiveness and training efficiency.

\subsubsection{$\mathbb{C}$GAL and $\mathbb{C}$FL in BeamGNN}

Two parallel branches are employed to separately learn $\{\alpha_{b,k}\}$ and $\{p_{b,k}\}$.

For hybrid coefficients, $G_4$ $\mathbb{C}$GAL layers followed by $G_5$ $\mathbb{C}$FL layers construct the mapping from updated features of BS-UE nodes to $\mathbf{V}_{G_5}^{\prime\prime}\in\mathbb{C}^{BK\times 1}$, and the sigmoid activation is applied:
\begin{flalign}
\alpha_{b,k} = \sigma\left(\mathrm{Re}\left([\mathbf{V}_{G_5}^{\prime\prime}]_{(b-1)K+k,:}\right)\right),~\forall b,k.
\end{flalign}

For power allocation, $G_6$ $\mathbb{C}$GAL layers followed by $G_7$ $\mathbb{C}$FL layers construct the mapping from node features to $\mathbf{V}_{G_7}^{\prime\prime}\in\mathbb{C}^{BK\times 1}$, yielding unconstrained power values:
\begin{flalign}
\widetilde{p}_{b,k} = P_{\max}\sigma\left(\mathrm{Re}\left([\mathbf{V}_{G_7}^{\prime\prime}]_{(b-1)K+k,:}\right)\right),~\forall b,k.
\end{flalign}

Unconstrained beamforming vectors $\{\widehat{\mathbf{w}}_{b,k}\}$ are then recovered following \eqref{HZM} with $\{\alpha_{b,k}\}$ and $\{\widetilde{p}_{b,k}\}$.

\subsection{Stage 3: AssocGNN}

The input of AssocGNN is the (unconstrained) beamforming gains computed with the obtained $\{\widehat{\mathbf{w}}_{b,k}\}$, and its  output is $\mathbf{U}$.

\subsubsection{Initial Node Features}
For BS-UE node $(b,k)$, its node feature is  initialized by the beamforming gain:
\begin{flalign}\label{beamgain}
[\mathbf{V}_0^{\prime}]_{(b-1)K+k,:} = \left|\widehat{\mathbf{h}}_{b,k}^H\widehat{\mathbf{w}}_{b,k}\right|^2,~\forall k,b.
\end{flalign}

\subsubsection{$\mathbb{C}$GAL and $\mathbb{C}$FL in AssocGNN}
$G_8$ $\mathbb{C}$GAL layers followed by $G_9$ $\mathbb{C}$FL layers map $\mathbf{V}_0^{\prime}$ to $\mathbf{V}_{G_9}^{\prime\prime}$, from which the association logits are extracted as
\begin{flalign}
\ell_{b,k} = \mathrm{Re}\left([\mathbf{V}_{G_9}^{\prime\prime}]_{(b-1)K+k,:}\right),~\forall k,b.
\end{flalign}

\subsubsection{Differentiable Association via Gumbel-Softmax}

To enable end-to-end differentiable training while satisfying constraints \eqref{p1:2f} and \eqref{p1:2g}, the Gumbel-Softmax \cite{Gumbel-Softmax} is employed. 

During the training stage, we set
\begin{flalign}\label{cons-u}
u_{b,k} = \frac{\exp\left(\left(\ell_{b,k}+g_{b,k}\right)/\tau_{\rm gs}\right)}{\sum_{b^{\prime}\in\mathcal{B}}\exp\left(\left(\ell_{b^{\prime},k}+g_{b^{\prime},k}\right)/\tau_{\rm gs}\right)},
\end{flalign}
where $g_{b,k}\sim\mathrm{Gumbel}(0,1)$ i.i.d. and $\tau_{\rm gs}>0$ is the temperature. During the inference stage, the hard assignment $u_{b,k} = \mathbf{1}[b=\arg\max_{b^{\prime}}\ell_{b^{\prime},k}]$ is used, satisfying \eqref{p1:2f} and \eqref{p1:2g} exactly.

\subsubsection{Feasibility-Guaranteed Power Allocation}

To enforce constraint \eqref{p1:2d} for BS $b$, a per-BS normalization is applied to $\widetilde{p}_{b,k}$:
\begin{flalign}\label{cons-p}
&p_{b,k} :=\\
& \left\{\begin{array}{ll}
\widetilde{p}_{b,k}, & \displaystyle\sum\limits_{k^{\prime}\in{\cal K}}u_{b,k^{\prime}}\widetilde{p}_{b,k^{\prime}}\le P_{\max}\\[4pt]
\dfrac{\widetilde{p}_{b,k}}{\sum_{k^{\prime}\in {\cal K}}u_{b,k^{\prime}}\widetilde{p}_{b,k^{\prime}}}P_{\max}, & \text{otherwise}
\end{array}\right.,~\forall b,\nonumber
\end{flalign}
such that $\sum_{k\in {\cal K}} u_{b,k}\|\mathbf{w}_{b,k}\|^2\le P_{\max}$ holds for all BSs.

Thus, the beamforming vectors $\{{\mathbf{w}}_{b,k}\}$ are recovered following \eqref{HZM} with $\{\alpha_{b,k}\}$ and $\{p_{b,k}\}$.

\subsection{Detailed Processes of ${\mathbb C}$HAL, ${\mathbb C}$GAL, and ${\mathbb C}$FL}

This subsection elaborates on the three layer types employed in our model, namely the $\mathbb{C}$HAL, $\mathbb{C}$GAL, and $\mathbb{C}$FL.

\subsubsection{Complex Heterogeneous Graph Attention Layer}

The ${\mathbb C}$HAL performs feature extraction via message passing over heterogeneous graph edges. To enrich representational capacity, a two-tier hierarchical attention mechanism is adopted: node-level attention captures the relative importance of neighboring nodes along a given meta-path, while semantic-level attention consolidates information across all meta-paths \cite{han}.

For the $g$-th ($g\in\{1,\ldots,G\}$) ${\mathbb C}$HAL, let ${\bf{V}}_g^{\Psi} \in {\mathbb C}^{M_\Psi \times S_g}$ denote the output feature matrix for nodes of type $\Psi\in\{\text{BS}, \text{UE}, \text{RIS}\}$, where $M_\Psi$ is the number of nodes of type $\Psi$ and $S_g$ is the per-node feature dimension. The input to the $g$-th ${\mathbb C}$HAL is $\{{\bf V}_{g-1}^{\Psi}\}_{\Psi}$, where the initial features ${\bf V}_{0}^{\rm BS}$, ${\bf V}_{0}^{\rm UE}$, and ${\bf V}_{0}^{\rm RIS}$ are exactly those defined in the Stage 1 input construction.

\textbf{Node-Level Attention.}
Each ${\mathbb C}$HAL employs $D$ parallel attention heads. For an ordered edge type connecting node types $\Psi$ and $\Phi$ ($\Psi,\Phi\in\{\text{BS}, \text{UE}, \text{RIS}\}$), the node-level attention coefficient matrix of the $d$-th head in the $g$-th layer, ${\bf A}^{\Psi,\Phi}_{g,d}\in{\mathbb R}^{M_\Psi\times M_\Phi}$, is computed via \eqref{node-level attention coefficient}, where ${\rm {\mathbb R}LeakyReLU}(\cdot)$ is the real-valued LeakyReLU activation, ${\bf W}_{g,d}^{\Psi}\in{\mathbb C}^{S_{g-1}\times S_g}$ is the feature transformation matrix, ${\bf a}_{g,d}^{\Psi,\Phi}\in{\mathbb C}^{2S_g}$ is the attention weight vector, and ${\cal N}_i^{\Psi,\Phi}$ denotes the set of type $\Phi$ neighbors of the $i$-th type $\Psi$ node.

\begin{figure*}
\begin{flalign}
\label{node-level attention coefficient}
{[{\bf{A}}_{g,d}^{{\Psi},{\Phi}}]_{i,j}} = \frac{{\exp\left({{\rm{{\mathbb R}LeakyReLU}}\left({{\rm Re}\left({{\bf{a}}_{g,d}^{{\Psi},{\Phi}}}^T {\rm Concat}\left({\left[{\bf{V}}_{g-1}^{\Psi}\right]_{i,:}}{\bf W}_{g,d}^{\Psi},\,{\left[{\bf{V}}_{g-1}^{\Phi}\right]_{j,:}}{\bf W}_{g,d}^{\Phi}\right)\right)}\right)}\right)}}
{{\sum\nolimits_{k\in{\cal N}_i^{\Psi,\Phi}}\exp\left({{\rm{{\mathbb R}LeakyReLU}}\left({{\rm Re}\left({{\bf{a}}_{g,d}^{{\Psi},{\Phi}}}^T {\rm Concat}\left({\left[{\bf{V}}_{g-1}^{\Psi}\right]_{i,:}}{\bf W}_{g,d}^{\Psi},\,{\left[{\bf{V}}_{g-1}^{\Phi}\right]_{k,:}}{\bf W}_{g,d}^{\Phi}\right)\right)}\right)}\right)}}
\end{flalign}
\hrule
\end{figure*}

The $d$-th head aggregates weighted type $\Phi$ neighbor features for the $i$-th type $\Psi$ node as 
\begin{flalign}\label{node-level update1}
{\bf x}_{g,i,d}^{\Psi,\Phi} = \mathbb{C}{\rm ReLU}\left(\sum_{j\in\mathcal{N}_{i}^{\Psi,\Phi}}[{\bf A}_{g,d}^{\Psi,\Phi}]_{i,j}[\mathbf{V}_{g-1}^{\Phi}]_{j,:}\mathbf{W}_{g,d}^{\Phi}\right),
\end{flalign}
where ${\rm \mathbb{C}ReLU}(\cdot)$ denotes the complex ReLU activation \cite{complex NN}. The $D$ head outputs are then concatenated to form the intermediate feature matrix ${\bf V}_g^{\Psi,\Phi}\in{\mathbb C}^{M_\Psi\times S_g}$:
\begin{flalign}\label{node-level update2}
{\left[{\bf{V}}_g^{{\Psi},{\Phi}}\right]_{i,:}} = {\rm Concat}\left({\bf x}_{g,i,1}^{{\Psi},{\Phi}},\,{\bf x}_{g,i,2}^{{\Psi},{\Phi}},\,\ldots,\,{\bf x}_{g,i,D}^{{\Psi},{\Phi}}\right).
\end{flalign}

\textbf{Semantic-Level Attention.}
To weight the contribution of each meta-path, semantic-level attention coefficients $B_g^{\Psi,\Phi}\in{\mathbb R}$ are computed as in \eqref{semantic-level attention coefficient}, where ${\rm \mathbb{R}tanh}(\cdot)$ is the real tanh activation, $\overline{\bf W}_g\in{\mathbb C}^{S_g\times S_g}$ and ${\bf q}_g\in{\mathbb C}^{S_g}$ are learnable parameters, ${\cal V}_\Psi$ is the set of type $\Psi$ nodes, and ${\cal N}_\Psi$ is the set of node types adjacent to type $\Psi$ nodes.

\begin{figure*}
\begin{flalign}
\label{semantic-level attention coefficient}
B_g^{{\Psi},{\Phi}} = \frac{{\exp\left({\frac{1}{|{\cal V}_\Psi|}\sum\nolimits_{i\in{\cal V}_\Psi}{\bf q}_g^T\cdot{\rm\mathbb{R}tanh}\left({\rm Re}\left({\left[{\bf V}_g^{\Psi,\Phi}\right]_{i,:}}\overline{\bf W}_g\right)\right)}\right)}}
{{\sum\nolimits_{\Phi^{\prime}\in{\cal N}_\Psi}\exp\left({\frac{1}{|{\cal V}_\Psi|}\sum\nolimits_{i\in{\cal V}_\Psi}{\bf q}_g^T\cdot{\rm\mathbb{R}tanh}\left({\rm Re}\left({\left[{\bf V}_g^{\Psi,\Phi^{\prime}}\right]_{i,:}}\overline{\bf W}_g\right)\right)}\right)}}
\end{flalign}
\hrule
\end{figure*}

The output ${\bf V}_g^\Psi$ is then obtained via weighted fusion across meta-paths with a residual connection to alleviate over-smoothing:
\begin{flalign}
\label{semantic-level update}
{\bf V}_g^\Psi = {\rm\mathbb{C}ReLU}\left(\sum\nolimits_{\Phi\in{\cal N}_\Psi}B_g^{\Psi,\Phi}{\bf V}_g^{\Psi,\Phi} + {\bf V}_{g-1}^\Psi\widehat{\bf W}_g^\Psi\right),
\end{flalign}
where $\widehat{\bf W}_g^\Psi\in{\mathbb C}^{S_{g-1}\times S_g}$ is the learnable residual matrix.

\subsubsection{Complex Graph Attention Layer}

The $\mathbb{C}$GAL captures inter-node interactions through the attention mechanism over a homogeneous graph, whereby each node assigns adaptive importance weights to its neighbors \cite{gat}.

For the $g$-th $\mathbb{C}$GAL with input features $\mathbf{V}^{\prime}_{g-1}\in\mathbb{C}^{BK\times S^{\prime}_{g-1}}$, the attention coefficient from node $k^{\prime}$ to node $k$ is given by \eqref{attention coefficient}, where $\mathbf{W}^{\prime}_g\in\mathbb{C}^{S^{\prime}_{g-1}\times S^{\prime}_g}$ and $\mathbf{a}_g\in\mathbb{C}^{2S^{\prime}_g}$ are learnable parameters, ${\rm\mathbb{R}LeakyReLU}(\cdot)$ denotes the real LeakyReLU activation, and $\mathcal{N}_k$ denotes the set of neighbors of node $k$.

\begin{figure*}
\begin{flalign}
\label{attention coefficient}
[{\bf A}_g]_{k,k^{\prime}} =
\frac{\exp\left({\rm\mathbb{R}LeakyReLU}\left({\rm Re}\left({\bf a}_g^T\,{\rm Concat}\left([\mathbf{V}^{\prime}_{g-1}]_{k,:}\mathbf{W}^{\prime}_g,\,[\mathbf{V}^{\prime}_{g-1}]_{k^{\prime},:}\mathbf{W}^{\prime}_g\right)\right)\right)\right)}
{\sum\nolimits_{k^{\prime\prime}\in\mathcal{N}_k}\exp\left({\rm\mathbb{R}LeakyReLU}\left({\rm Re}\left({\bf a}_g^T\,{\rm Concat}\left([\mathbf{V}^{\prime}_{g-1}]_{k,:}\mathbf{W}^{\prime}_g,\,[\mathbf{V}^{\prime}_{g-1}]_{k^{\prime\prime},:}\mathbf{W}^{\prime}_g\right)\right)\right)\right)}
\end{flalign}
\hrule
\end{figure*}

The updated node features are produced by aggregating attention-weighted neighbor representations:
\begin{flalign}
\left[\mathbf{V}^{\prime}_g\right]_{k,:} = \mathbb{C}{\rm ReLU}\left(\sum\nolimits_{k^{\prime}\in\mathcal{N}_k}[{\bf A}_g]_{k,k^{\prime}}\left[\mathbf{V}^{\prime}_{g-1}\right]_{k^{\prime},:}\mathbf{W}^{\prime}_g\right).
\end{flalign}

A residual connection is introduced to stabilize training and mitigate over-smoothing:
\begin{flalign}
\left[\mathbf{V}^{\prime}_g\right]_{k,:}  := \left[\mathbf{V}^{\prime}_g\right]_{k,:} + \left[\mathbf{V}^{\prime}_{g-1}\right]_{k,:}\overline{\mathbf{W}}^{\prime}_g + \left[\mathbf{V}^{\prime}_{0}\right]_{k,:}\widehat{\mathbf{W}}^{\prime}_g,
\end{flalign}
where $\overline{\mathbf{W}}^{\prime}_g\in\mathbb{C}^{S^{\prime}_{g-1}\times S^{\prime}_g}$ and $\widehat{\mathbf{W}}^{\prime}_g\in\mathbb{C}^{S^{\prime}_{0}\times S^{\prime}_g}$ are learnable residual matrices.

\subsubsection{Complex Fully-Connected Layer}

The $\mathbb{C}$FL maps the embedding produced by the $\mathbb{C}$HALs or $\mathbb{C}$GALs to the target vector via feedforward operations adapted for complex-valued data.

For the $f$-th $\mathbb{C}$FL with input $\mathbf{V}^{\prime\prime}_{f-1}\in\mathbb{C}^{M_\Psi\times S^{\prime\prime}_{f-1}}$, the output is computed as
\begin{flalign}
\mathbf{V}^{\prime\prime}_f = \mathbb{C}{\rm ReLU}\left(\mathbf{V}^{\prime\prime}_{f-1}\mathbf{W}^{\prime\prime}_f + \mathbf{B}_f\right),
\end{flalign}
where $\mathbf{W}^{\prime\prime}_f\in\mathbb{C}^{S^{\prime\prime}_{f-1}\times S^{\prime\prime}_f}$ and $\mathbf{B}_f\in\mathbb{C}^{M_\Psi\times S^{\prime\prime}_f}$ are learnable weight matrices. Each $\mathbb{C}$FL is followed by a complex batch normalization layer \cite{complex NN} to improve convergence and reduce overfitting.

\subsection{Unsupervised Loss Function}

Given input node location features, the proposed GNN sequentially produces the complete solution $\{x_{b,n,m}^{\rm P}, \{\bm{\Phi}_r\}, \{\mathbf{w}_{b,k}\}, \mathbf{U}\}$, enabling end-to-end unsupervised training by directly optimizing system utility. 

\begin{table*}[t]
\centering
\caption{Constraint enforcement mechanisms in the proposed three-stage GNN.}
\label{tab:constraint_enforcement}
\renewcommand{\arraystretch}{1.15}
\setlength{\tabcolsep}{4.5pt}
\begin{tabular}{p{4.9cm} p{1.1cm} p{8.2cm} p{2.3cm}}
\toprule
\textbf{Constraint in \eqref{p1}} & \textbf{Stage} & \textbf{Enforcement mechanism} & \textbf{Key equations} \\
\midrule
\eqref{p1:2b} $0 \le x_{b,n,m}^{P} \le C_b$ 
& Stage 1 
& Re-parameterize PA positions by auxiliary spacing variables $\delta_{b,n,m}$, apply sigmoid scaling and per-waveguide normalization, and then reconstruct $x_{b,n,m}^{P}$ 
& \eqref{delta}--\eqref{sum_x} \\

\eqref{p1:2c} $x_{b,n,m}^{P} - x_{b,n,m-1}^{P} \ge \Delta_{\min}$ 
& Stage 1 
& Same spacing-based feasibility-preserving reconstruction as above 
& \eqref{delta}--\eqref{sum_x} \\

\eqref{p1:2d} $\sum_{k\in\mathcal{K}} u_{b,k}\|w_{b,k}\|^2 \le P_{\max}$ 
& Stage 3 
& Apply per-BS power normalization after association is obtained 
& \eqref{cons-p} \\

\eqref{p1:2e} $\phi_{r,l}\in[0,2\pi)$ 
& Stage 1 
& Normalize each complex output to unit modulus and construct $\Phi_r$ accordingly 
& \eqref{phi_init}, \eqref{cons-phi} \\

\eqref{p1:2f} $\sum_{b\in\mathcal{B}} u_{b,k}=1$ 
& Stage 3 
& Use Gumbel-Softmax during training and hard one-hot assignment during inference 
& \eqref{cons-u} \\

\eqref{p1:2g} $u_{b,k}\in\{0,1\}$ 
& Stage 3 
& Use argmax-based hard assignment at inference; differentiable relaxation is used only for training 
& \eqref{cons-u} \\
\bottomrule
\end{tabular}
\end{table*}

Denote all learnable parameters as $\bm{\Theta}$, encompassing the parameters of all $\mathbb{C}$HALs, $\mathbb{C}$GALs, and $\mathbb{C}$FLs.  The proposed GNN solves both $\rm P1$ and $\rm P2$ in a unified framework, differing only in the loss function. For Problem $\rm P1$:
\begin{flalign}\label{loss1}
&\mathcal{L}_T(\bm{\Theta}) = \notag\\
&\frac{1}{T}\sum_{t=1}^T\frac{1}{\sum_{k\in\cal{K}}R_k^{(t)}\left(\left\{x_{b,n,m}^{\rm P},\{\bm{\Phi}_r\},\{\mathbf{w}_{b,k}\},\mathbf{U}\,\big|\,\bm{\Theta}\right\}\right)}.
\end{flalign}
For Problem P2:
\begin{flalign}\label{loss2}
&\mathcal{L}_T(\bm{\Theta}) = \notag\\
&\frac{1}{T}\sum_{t=1}^T\frac{\sum_{b\in\mathcal{B}}\sum_{k\in\cal{K}}u_{b,k}^{(t)}\|\mathbf{w}_{b,k}^{(t)}\|^2+P_{\rm C}}{\sum_{k\in\cal{K}}R_k^{(t)}\left(\left\{x_{b,n,m}^{\rm P},\{\bm{\Phi}_r\},\{\mathbf{w}_{b,k}\},\mathbf{U}\,\big|\,\bm{\Theta}\right\}\right)}.
\end{flalign}

Notably, there are no penalty terms associated with constraints in \eqref{loss1} and \eqref{loss2}, as all constraints are guaranteed to be satisfied. As summarized in Table~\ref{tab:constraint_enforcement}, the constraints in \eqref{p1:2b}–\eqref{p1:2g} are enforced by feasibility-preserving mappings rather than by soft penalties.


\begin{rem}\label{sca} (Scalability with the numbers of UEs, BSs and RISs)
Among the learnable parameters in $\bm \Theta$, only the bias term ${\bf B}_f$, is nominally dependent on the numbers of UEs, BSs, and RISs through its row dimension. Nevertheless, the ${M_{\Psi} }$ row vectors of ${\bf B}_f$ can be set identically for any value of ${M_{\Psi}}$. Hence, all computations in the proposed GNN can be implemented independently of ${M_{\Psi}}$. By parameter sharing, proposed GNN is scalable to the numbers of UEs, BSs and RISs, ensuring that proposed GNN is acceptable to unseen problem sizes during both training and test phases.
\end{rem}

\section{Numerical Results}

This section provides numerical results to evaluate the proposed three-stage GNN. 

\subsection{Simulation Setting}

\subsubsection{Simulation Scenario}

The simulation scenario follows Figure~\ref{sys}. We consider $B\in\{1,2,3,4\}$ PA-BSs and $R\in\{1,2,3,4\}$ RISs deployed to serve $K\in\{2,\ldots,6,10,\ldots,18\}$ single-antenna UEs within a rectangular region of length $D$ m and width $S$ m. Each BS is equipped with $N\in\{8,16,20\}$ parallel waveguides each hosting $M\in\{2,3,4,5,6\}$ movable PAs, and each RIS comprises $L\in\{16,64\}$ reflecting elements. The total power budget is $P_{\max}=10$ W, the circuit power is $P_{\rm C}=5$ W, and the noise power is $\sigma_k^2=-60$ dBm. The simulation parameters are summarized in Table~\ref{Simulation Parameters}.

UEs are distributed uniformly within $[0,D]\times[0,S]$. The $B$ BSs are placed on a uniform grid within the same region, with each BS's $N$ waveguides offset evenly in the $y$-direction with spacing $\Delta_{\rm wg}$ and placed at height $H_{\rm b}$, with feed points at the left end of each waveguide. The $R$ RISs are similarly placed on a uniform grid at height $H_{\rm b}/2$. Specifically, given $B$ BSs (or $R$ RISs), the grid dimensions $N_{\rm row}$ and $N_{\rm col}$ are jointly determined by solving
\begin{flalign}
(N_{\rm row}, N_{\rm col}) = \mathop{\arg\min}_{p,q:\,pq \ge B,\,p,q\in\mathbb{Z}^+} 
\left|\frac{q}{p} - \frac{D}{S}\right|,
\end{flalign}
so that the grid aspect ratio $N_{\rm col}/N_{\rm row}$ best matches the region aspect ratio $D/S$. The grid-center coordinates are then assigned as
\begin{flalign}
x_c = \left(c - \tfrac{1}{2}\right)\frac{D}{N_{\rm col}},\quad 
y_r = \left(r - \tfrac{1}{2}\right)\frac{S}{N_{\rm row}},
\end{flalign}
where $c\in\{1,\ldots,N_{\rm col}\}$ and $r\in\{1,\ldots,N_{\rm row}\}$ index the grid columns and rows, respectively.

\begin{table}[t]
\footnotesize
\centering
\caption{Simulation Parameters.}
\label{Simulation Parameters}
\begin{tabular}{c|c}
\hline
{\bf Parameter} & {\bf Value} \\ \hline\hline
Number of BSs           & $B\in\{1,2,3,4\}$ \\ \hline
Number of RISs          & $R\in\{1,2,3,4\}$ \\ \hline
Number of waveguides per BS & $N\in\{8,16,20\}$ \\ \hline
Number of PAs per waveguide & $M\in\{2,3,4,5,6\}$ \\ \hline
Number of UEs         & $K\in\{2,\ldots,6,10,\ldots,18\}$ \\ \hline
Number of RIS elements  & $L\in\{16,64\}$ \\ \hline
Serving area            & $S=D\in\{30,40,50,60,70\}$ m\\ \hline
Waveguide height        & $H_{\rm b}=5$  m \\ \hline
Waveguide length        & $C=10$  m \\ \hline
Waveguide $y$-spacing   & $\Delta_{\rm wg}=0.7$  m \\ \hline
Power budget            & $P_{\max}=10$ W \\ \hline
Circuit power           & $P_{\rm C}=5$ W \\ \hline
Noise power             & $\sigma_k^2=-60$ dBm \\ \hline
Carrier frequency       & $f_c=6$ GHz \\ \hline
Free-space wavelength   & $\lambda=c/f_c=0.05$ m \\ \hline
Minimum PA spacing  & $\Delta_{\min}=0.1$ m \\ \hline
Effective refractive index & $n_{\rm eff}=1.4$ \\ \hline
In-waveguide attenuation & $\zeta=0.0046$  \\ \hline
Rician factor           & $\kappa=3$ dB \\ \hline
Path loss exponent      & $\alpha=2.8$ \\ \hline
Channel gain at 1 m     & $\beta_0=-20$ dB \\ \hline
\end{tabular}
\end{table}

\subsubsection{Baselines}

The proposed system with jointly optimized PA positions, RIS phase shifts, beamforming vectors, and BS-UE association is denoted \emph{Proposed GNN}. To evaluate the contribution of each component, the following baselines are considered.

\textbf{System baselines:}
\begin{itemize}
    \item \emph{No-RIS PA}: The case without RIS assistance, where $\{\bm{\Phi}_r\}$ is removed and the remaining other variables are jointly learned.
    \item \emph{Fixed-PA}: The case where PA positions are fixed at equal spacing $\Delta_{\min}$ on each waveguide, with $\{\bm{\Phi}_r\}$, $\{\mathbf{w}_{b,k}\}$, and $\mathbf{U}$ jointly learned.
    \item \emph{No-RIS Fixed-PA}: The case without RIS assistance and with fixed PA positions, serving as a conventional fixed-position antenna baseline.
    \item \emph{Random-$\mathbf{U}$}: The BS-UE association matrix $\mathbf{U}$ is generated randomly at each inference, while $\{x_{b,n,m}^{\rm P}\}$, $\{\bm{\Phi}_r\}$, and $\{\mathbf{w}_{b,k}\}$ are jointly learned, isolating the gain from association optimization.
\end{itemize}

\textbf{Model baselines:}
\begin{itemize}
    \item \emph{MLP}: A basic feedforward MLP with HZM learning that directly maps node location features to the complete solution, without exploiting graph structure.
    \item \emph{Single HAN} \cite{han}: A single-stage complex heterogeneous graph attention network (HAN) with HZM learning, in which the UE nodes output $\{\mathbf{w}_{b,k}\}$ and $\mathbf{U}$, the RIS nodes output $\{\bm{\Phi}_r\}$, and the BS nodes output $\{x_{b,n,m}^{\rm P}\}$.
    \item \emph{HAN}: A three-stage complex HAN with HZM learning, using the same node output design for UE, RIS, and BS nodes.
    \item \emph{GAT} \cite{gat}: A four-stage complex GAT with HZM learning, in which the system is modeled as a fully-connected homogeneous graph with $K$ nodes. The input includes only UE positions, while the outputs $\{\bm{\Phi}_r\}$ and $\{x_{b,n,m}^{\rm P}\}$ are obtained by averaging the embedding over all nodes \cite{rispassgnn2025}.
\end{itemize}

\subsubsection{Training and Dataset}

All learnable parameters are initialized using the Kaiming normal initialization method~\cite{kaiming_normal} with an initial learning rate of $5\times10^{-5}$. A multi-step learning rate scheduler adaptively reduces the learning rate during training. The Adam optimizer~\cite{Adam} is used for gradient-based updates over $50$ epochs with a batch size of $128$. An early stopping mechanism monitors validation performance and retains the parameter set yielding the best validation metric.

Each training sample is generated by independently drawing $K$ UE locations uniformly at random within the rectangular region, while BS and RIS locations are fixed across all samples according to the uniform grid placement described above. The NLoS fading components $\mathbf{l}_{b,r}^{\rm NLoS}\sim\mathcal{CN}(\mathbf{0},\mathbf{I}_L)$ and $\mathbf{h}_{r,k}^{\rm NLoS}\sim\mathcal{CN}(\mathbf{0},\mathbf{I}_L)$ are independently redrawn for each sample. Two data categories are constructed: a primary dataset of $100,000$ samples split into training, validation, and test subsets in an 8:1:1 ratio, and a generalization dataset of $10,000$ samples with unseen system configurations used exclusively for testing.

\subsubsection{Computer Configuration}

All models are trained and tested under Python 3.10 with PyTorch 1.11.0 on a computer equipped with an Intel Xeon Gold 6278C CPU and an NVIDIA Tesla V100 GPU (32 GB memory).

\subsection{Performance Comparison under Varying Problem Sizes}

\begin{table*}[t]
\belowrulesep=-0.3pt
\aboverulesep=-0.3pt
\centering
\caption{Performance comparison of different system baselines under varying problem sizes.}
\label{tab:system_baseline_comparison}
\renewcommand{\arraystretch}{1.15}
\setlength{\tabcolsep}{4pt}
\begin{tabular}{ccccccc||cc|cc|cc|cc}
\toprule
\multirow{2}{*}{\makebox[0.025\textwidth]{\textbf{$B$}}} & \multirow{2}{*}{\textbf{\makebox[0.025\textwidth]{$R$}}} & \multirow{2}{*}{\textbf{\makebox[0.025\textwidth]{$N$}}} & \multirow{2}{*}{\textbf{\makebox[0.025\textwidth]{$L$}}} & \multirow{2}{*}{\textbf{\makebox[0.025\textwidth]{$M$}}} & \multirow{2}{*}{\textbf{\makebox[0.025\textwidth]{$K_{\rm tr}$}}} & \multirow{2}{*}{\makebox[0.025\textwidth]{\textbf{$K_{\rm te}$}}} & \multicolumn{2}{c|}{\textbf{Proposed GNN}} & \multicolumn{2}{c|}{\textbf{No-RIS PA}} & \multicolumn{2}{c|}{\textbf{No-RIS Fixed-PA}} & \multicolumn{2}{c}{\textbf{Random-$\mathbf{U}$}} \\
\cmidrule(lr){8-9}\cmidrule(lr){10-11}\cmidrule(lr){12-13}\cmidrule(lr){14-15}
& & & & & & & \makebox[0.072\textwidth]{EE} & \makebox[0.072\textwidth]{SR} & \makebox[0.072\textwidth]{EE} & \makebox[0.072\textwidth]{SR} & \makebox[0.072\textwidth]{EE} & \makebox[0.072\textwidth]{SR} & \makebox[0.072\textwidth]{EE} & \makebox[0.072\textwidth]{SR} \\
\midrule
\multirow{5}{*}{1} & \multirow{5}{*}{1} & \multirow{5}{*}{8} & \multirow{5}{*}{64} & \multirow{5}{*}{6} & \multirow{5}{*}{4} & 2 & 4.12 & 31.55 & 3.62 & 28.72 & 3.12 & 25.60 & 4.12 & 31.55 \\
& & & & & & 3 & 5.81 & 44.69 & 5.10 & 40.51 & 4.27 & 35.55 & 5.81 & 44.69 \\
& & & & & & 4 & \cellcolor{blue!13}7.29 & \cellcolor{blue!13}56.74 & \cellcolor{blue!13}6.34 & \cellcolor{blue!13}51.27 & \cellcolor{blue!13}5.22 & \cellcolor{blue!13}44.44 & \cellcolor{blue!13}7.29 & \cellcolor{blue!13}56.74 \\
& & & & & & 5 & 8.48 & 67.28 & 7.33 & 60.67 & 5.95 & 52.04 & 8.48 & 67.28 \\
& & & & & & 6 & 9.38 & 76.16 & 8.03 & 68.31 & 6.45 & 58.23 & 9.38 & 76.16 \\
\midrule
\multicolumn{7}{c||}{Inference time} & \multicolumn{2}{c|}{0.81 ms} & \multicolumn{2}{c|}{0.76 ms} & \multicolumn{2}{c|}{0.57 ms} & \multicolumn{2}{c}{0.72 ms} \\
\midrule
\multirow{5}{*}{4} & \multirow{5}{*}{4} & \multirow{5}{*}{16} & \multirow{5}{*}{16} & \multirow{5}{*}{6} & \multirow{5}{*}{12} & 10 & 17.10 & 155.92 & 15.98 & 149.02 & 13.66 & 143.13 & 14.31 & 127.49 \\
& & & & & & 11 & 18.10 & 167.17 & 16.89 & 160.10 & 14.11 & 153.21 & 15.04 & 136.10 \\
& & & & & & 12 & \cellcolor{blue!13}18.84 & \cellcolor{blue!13}177.59 & \cellcolor{blue!13}17.61 & \cellcolor{blue!13}170.02 & \cellcolor{blue!13}14.44 & \cellcolor{blue!13}162.25 & \cellcolor{blue!13}15.54 & \cellcolor{blue!13}143.78 \\
& & & & & & 13 & 19.35 & 186.41 & 17.91 & 178.32 & 14.48 & 169.95 & 15.84 & 149.59 \\
& & & & & & 14 & 19.30 & 193.07 & 17.83 & 184.09 & 14.06 & 175.26 & 15.74 & 153.19 \\
\midrule
\multicolumn{7}{c||}{Inference time} & \multicolumn{2}{c|}{2.00 ms} & \multicolumn{2}{c|}{1.97 ms} & \multicolumn{2}{c|}{1.73 ms} & \multicolumn{2}{c}{1.64 ms} \\
\midrule
\multirow{5}{*}{4} & \multirow{5}{*}{4} & \multirow{5}{*}{20} & \multirow{5}{*}{16} & \multirow{5}{*}{6} & \multirow{5}{*}{16} & 14 & 22.41 & 209.95 & 21.12 & 201.96 & 18.03 & 191.69 & 18.75 & 184.54 \\
& & & & & & 15 & 23.07 & 219.80 & 21.71 & 211.38 & 18.36 & 199.86 & 19.23 & 192.70 \\
& & & & & & 16 & \cellcolor{blue!13}23.54 & \cellcolor{blue!13}228.47 & \cellcolor{blue!13}22.09 & \cellcolor{blue!13}219.46 & \cellcolor{blue!13}18.45 & \cellcolor{blue!13}207.47 & \cellcolor{blue!13}19.52 & \cellcolor{blue!13}199.57 \\
& & & & & & 17 & 23.63 & 239.75 & 22.17 & 229.55 & 18.13 & 213.57 & 19.48 & 203.99 \\
& & & & & & 18 & 23.17 & 234.11 & 21.63 & 224.01 & 17.36 & 217.09 & 18.97 & 205.26 \\
\midrule
\multicolumn{7}{c||}{Inference time} & \multicolumn{2}{c|}{2.56 ms} & \multicolumn{2}{c|}{2.54 ms} & \multicolumn{2}{c|}{2.28 ms} & \multicolumn{2}{c}{2.07 ms} \\
\bottomrule
\end{tabular}
\begin{tablenotes}
\footnotesize
\item $K_{\rm {Tr}}$/$K_{\rm {Te}}$: Value of $K$ in the training/test set.
\end{tablenotes}
\end{table*}

\begin{table*}[t]
\belowrulesep=-0.3pt
\aboverulesep=-0.3pt
\centering
\caption{Performance comparison of different model baselines under varying problem sizes.}
\label{tab:model_baseline_comparison}
\renewcommand{\arraystretch}{1.15}
\setlength{\tabcolsep}{4pt}
\begin{tabular}{ccccccc||cc|cc|cc|cc|cc}
\toprule
\multirow{2}{*}{\makebox[0.025\textwidth]{\textbf{$B$}}} 
& \multirow{2}{*}{\textbf{\makebox[0.025\textwidth]{$R$}}} 
& \multirow{2}{*}{\textbf{\makebox[0.025\textwidth]{$N$}}} 
& \multirow{2}{*}{\textbf{\makebox[0.025\textwidth]{$L$}}} 
& \multirow{2}{*}{\textbf{\makebox[0.025\textwidth]{$M$}}} 
& \multirow{2}{*}{\textbf{\makebox[0.025\textwidth]{$K_{\rm tr}$}}} 
& \multirow{2}{*}{\makebox[0.025\textwidth]{\textbf{$K_{\rm te}$}}} 
& \multicolumn{2}{c|}{\textbf{Proposed GNN}} 
& \multicolumn{2}{c|}{\textbf{MLP}} 
& \multicolumn{2}{c|}{\textbf{Single HAN}} 
& \multicolumn{2}{c|}{\textbf{HAN}} 
& \multicolumn{2}{c}{\textbf{GAT}} \\
\cmidrule(lr){8-9}\cmidrule(lr){10-11}\cmidrule(lr){12-13}\cmidrule(lr){14-15}\cmidrule(lr){16-17}
& & & & & & 
& \makebox[0.055\textwidth]{EE} & \makebox[0.055\textwidth]{SR} 
& \makebox[0.055\textwidth]{EE} & \makebox[0.055\textwidth]{SR} 
& \makebox[0.055\textwidth]{EE} & \makebox[0.055\textwidth]{SR} 
& \makebox[0.055\textwidth]{EE} & \makebox[0.055\textwidth]{SR} 
& \makebox[0.055\textwidth]{EE} & \makebox[0.055\textwidth]{SR} \\
\midrule
\multirow{5}{*}{1} 
& \multirow{5}{*}{1} 
& \multirow{5}{*}{8} 
& \multirow{5}{*}{64} 
& \multirow{5}{*}{6} 
& \multirow{5}{*}{4} 
& 2  & 4.12 & 31.55 & $\times$ & $\times$ & 2.10 & 31.51 & 2.12 & 31.51 & 4.12 & 31.51 \\
& & & & & & 3  & 5.81 & 44.69 & $\times$ & $\times$ & 2.97 & 44.69 & 3.00 & 44.75 & 5.80 & 44.65 \\
& & & & & & 4  & \cellcolor{blue!13}7.29 & \cellcolor{blue!13}56.74 & \cellcolor{blue!13}3.85 & \cellcolor{blue!13}43.14 & \cellcolor{blue!13}3.78 & \cellcolor{blue!13}56.58 & \cellcolor{blue!13}3.81 & \cellcolor{blue!13}56.70 & \cellcolor{blue!13}7.25 & \cellcolor{blue!13}56.70 \\
& & & & & & 5  & 8.48 & 67.28 & $\times$ & $\times$ & 4.48 & 67.30 & 4.52 & 67.37 & 8.45 & 67.26 \\
& & & & & & 6  & 9.38 & 76.16 & $\times$ & $\times$ & 5.09 & 76.32 & 5.12 & 76.38 & 9.35 & 76.26 \\
\midrule
\multicolumn{7}{c||}{Inference time} 
& \multicolumn{2}{c|}{0.81 ms} 
& \multicolumn{2}{c|}{0.03 ms} 
& \multicolumn{2}{c|}{0.68 ms} 
& \multicolumn{2}{c|}{1.19 ms} 
& \multicolumn{2}{c}{0.43 ms} \\
\midrule
\multirow{5}{*}{4} 
& \multirow{5}{*}{4} 
& \multirow{5}{*}{16} 
& \multirow{5}{*}{16} 
& \multirow{5}{*}{6} 
& \multirow{5}{*}{12} 
& 10 & 17.10 & 155.92 & $\times$ & $\times$ & 13.14 & 142.95 & 14.27 & 121.18 & 14.26 & 146.52 \\
& & & & & & 11 & 18.10 & 167.17 & $\times$ & $\times$ & 13.80 & 153.19 & 15.01 & 128.67 & 15.02 & 156.45 \\
& & & & & & 12 & \cellcolor{blue!13}18.84 & \cellcolor{blue!13}177.59 & \cellcolor{blue!13}8.68 & \cellcolor{blue!13}154.35 & \cellcolor{blue!13}14.16 & \cellcolor{blue!13}162.29 & \cellcolor{blue!13}15.52 & \cellcolor{blue!13}135.43 & \cellcolor{blue!13}15.52 & \cellcolor{blue!13}166.29 \\
& & & & & & 13 & 19.35 & 186.41 & $\times$ & $\times$ & 14.37 & 169.85 & 15.81 & 140.45 & 15.78 & 174.07 \\
& & & & & & 14 & 19.30 & 193.07 & $\times$ & $\times$ & 14.17 & 174.83 & 15.66 & 142.81 & 15.66 & 179.03 \\
\midrule
\multicolumn{7}{c||}{Inference time} 
& \multicolumn{2}{c|}{2.00 ms} 
& \multicolumn{2}{c|}{0.21 ms} 
& \multicolumn{2}{c|}{0.83 ms} 
& \multicolumn{2}{c|}{1.43 ms} 
& \multicolumn{2}{c}{0.83 ms} \\
\midrule
\multirow{5}{*}{4} 
& \multirow{5}{*}{4} 
& \multirow{5}{*}{20} 
& \multirow{5}{*}{16} 
& \multirow{5}{*}{6} 
& \multirow{5}{*}{16} 
& 14 & 22.41 & 209.95 & $\times$ & $\times$ & 18.18 & 196.00 & 18.76 & 160.50 & 18.67 & 196.67 \\
& & & & & & 15 & 23.07 & 219.80 & $\times$ & $\times$ & 18.61 & 204.72 & 19.29 & 166.19 & 19.14 & 205.27 \\
& & & & & & 16 & \cellcolor{blue!13}23.54 & \cellcolor{blue!13}228.47 & \cellcolor{blue!13}11.14 & \cellcolor{blue!13}200.26 & \cellcolor{blue!13}18.83 & \cellcolor{blue!13}212.44 & \cellcolor{blue!13}19.50 & \cellcolor{blue!13}171.47 & \cellcolor{blue!13}19.41 & \cellcolor{blue!13}212.83 \\
& & & & & & 17 & 23.63 & 239.75 & $\times$ & $\times$ & 18.79 & 217.90 & 19.44 & 173.84 & 19.37 & 218.43 \\
& & & & & & 18 & 23.17 & 234.11 & $\times$ & $\times$ & 18.38 & 220.78 & 18.99 & 173.68 & 18.84 & 221.70 \\
\midrule
\multicolumn{7}{c||}{Inference time} 
& \multicolumn{2}{c|}{2.56 ms} 
& \multicolumn{2}{c|}{0.25 ms} 
& \multicolumn{2}{c|}{0.91 ms} 
& \multicolumn{2}{c|}{1.58 ms} 
& \multicolumn{2}{c}{1.10 ms} \\
\bottomrule
\end{tabular}
\begin{tablenotes}
\footnotesize
\item $\times$ represents ``not applicable".
\end{tablenotes}
\end{table*}

Tables~\ref{tab:system_baseline_comparison} and~\ref{tab:model_baseline_comparison} compare different methods under varying problem sizes and mismatched training/test UE numbers. The proposed GNN consistently delivers the strongest overall EE and SR performance, which verifies the effectiveness of the proposed graph construction and stage-wise decomposition in jointly coordinating PA placement, RIS configuration, active beamforming, and BS-UE association. 

Table~\ref{tab:system_baseline_comparison} further demonstrates the effectiveness of the individual system components. Comparing the proposed GNN with No-RIS PA and No-RIS Fixed-PA shows that RIS reconfiguration and PA mobility are complementary. RISs provide environment-side controllability, whereas movable PAs improve transmitter-side spatial matching to the UE geometry. Once either component is removed, the performance degrades; when both are removed, the degradation becomes more evident. This confirms that the advantage of the proposed framework lies in jointly exploiting both types of reconfigurability rather than relying on only one of them. The Random-$\mathbf{U}$ baseline is also informative. In the single-BS case, it essentially coincides with the proposed GNN since association degenerates to a trivial decision. By contrast, in the multi-BS case, random association leads to a clear performance loss, which confirms that the proposed model effectively matches each UE to a more favorable serving BS.

Table~\ref{tab:model_baseline_comparison} validates the effectiveness of the model design. The significant performance gain of the GNN methods over the MLP indicates that the data contain rich graph-topological information, which can be effectively exploited by GNN-based models. In the single-BS single-RIS scenario, GAT achieves performance comparable to that of the proposed GNN, showing that the proposed GNN is at least as effective as architectures specifically tailored to homogeneous-graph scenarios. In the multi-BS multi-RIS setting, however, neither GAT based on homogeneous graph representation nor HAN based on heterogeneous graph representation can match the performance of the proposed GNN. This is because the homogeneous graph representation adopted by GAT cannot effectively exploit the heterogeneous information among the three node types from the outset, while HAN, after leveraging heterogeneous information, still struggles to distinguish inter-BS competition and inter-UE interference. These results confirm the effectiveness of the proposed GNN in jointly incorporating both heterogeneous and homogeneous graph representations.

Another important observation from Tables~\ref{tab:system_baseline_comparison} and~\ref{tab:model_baseline_comparison} is that the proposed GNN exhibits scalability with respect to the number of UEs, which alleviates the burden of retraining deep learning models to accommodate dynamic wireless environments during deployment. Specifically, both EE and SR increase approximately linearly with the number of UEs, while exhibiting a tendency toward saturation. Furthermore, the scalability becomes more limited as the problem size grows larger. In addition, the inference times remain at the millisecond level as the problem size grows, showing that the proposed GNN is suitable for online deployment with limited latency budget.

\subsection{Scalability with Respect to the Numbers of BSs and RISs}

\begin{table}[t]
\belowrulesep=-0.3pt
\aboverulesep=-0.3pt
\centering
\caption{Scalability with respect to the numbers of BSs and RISs.}
\label{tab:br_scalability}
\renewcommand{\arraystretch}{1.15}
\setlength{\tabcolsep}{6pt}

\textbf{EE (bit/J/Hz)}\\[2pt]
\begin{tabular}{c|cccc}
\toprule
\makebox[0.132\textwidth]{\diagbox[width=2.85cm]{$(B_{\rm {Tr}},R_{\rm {Tr}})$}{\textbf{$(B_{\rm {Te}},R_{\rm {Te}})$}}} & \makebox[0.055\textwidth]{\textbf{$(1,1)$}} & \makebox[0.055\textwidth]{\textbf{$(2,2)$}} & \makebox[0.055\textwidth]{\textbf{$(3,3)$}} & \makebox[0.055\textwidth]{\textbf{$(4,4)$}} \\
\midrule
\textbf{$(1,1)$} & \cellcolor{blue!13}11.23 & 11.38 & 11.43 & 11.48 \\
\textbf{$(2,2)$} & 9.46 & \cellcolor{blue!10}20.35 & 18.50 & 18.23 \\
\textbf{$(3,3)$} & 2.36 & 14.99 & \cellcolor{blue!10}21.61 & 21.67 \\
\textbf{$(4,4)$} & 1.86 &  2.80 & 14.25 & \cellcolor{blue!10}23.54 \\
\bottomrule
\end{tabular}

\vspace{0.6em}

\textbf{SR (bit/s/Hz)}\\[2pt]
\begin{tabular}{c|cccc}
\toprule
\makebox[0.132\textwidth]{\diagbox[width=2.85cm]{$(B_{\rm {Tr}},R_{\rm {Tr}})$}{{$(B_{\rm {Te}},R_{\rm {Te}})$}}} & \makebox[0.055\textwidth]{\textbf{$(1,1)$}} & \makebox[0.055\textwidth]{\textbf{$(2,2)$}} & \makebox[0.055\textwidth]{\textbf{$(3,3)$}} & \makebox[0.055\textwidth]{\textbf{$(4,4)$}} \\
\midrule
\textbf{$(1,1)$} & \cellcolor{blue!10}168.42 & 184.59 & 194.55 & 199.61 \\
\textbf{$(2,2)$} & 113.59 & \cellcolor{blue!10}194.13 & 214.50 & 227.05 \\
\textbf{$(3,3)$} & 106.15 & 181.99 & \cellcolor{blue!10}215.10 & 227.91 \\
\textbf{$(4,4)$} & 108.16 & 159.85 & 210.08 & \cellcolor{blue!10}228.47 \\
\bottomrule
\end{tabular}
\begin{tablenotes}
\footnotesize
\item $(B_{\rm {Tr}},R_{\rm {Tr}})$/$(B_{\rm {Te}},R_{\rm {Te}})$: Value of $B$ and $R$ in the training/test set.
\end{tablenotes}
\end{table}

Table~\ref{tab:br_scalability} evaluates the scalability of the proposed GNN across different numbers of BSs and RISs. The diagonal entries achieve the best performance, as the training and test graph topologies are matched. Nevertheless, the off-diagonal results show that the proposed GNN still exhibits scalability to unseen $(B,R)$ settings, especially when the topology mismatch is moderate. This demonstrates that the proposed architecture captures reusable structural relations among BSs, RISs, and UEs, rather than memorizing only one specific deployment scenario. 

At the same time, Table~\ref{tab:br_scalability} also reveals that scalability across $(B,R)$ is more challenging than generalization across the UE number $K$. Changing the numbers of BSs and RISs alters not only the node count, but also the coordination topology, the interference pattern, and the feasible association space. Therefore, large topology mismatches lead to more visible performance degradation, which is particularly pronounced for EE.

\subsection{Architectural Ablation}

\begin{table}[t]
\centering
\caption{Ablation experiment with $(B,R,N,L,M,K,S,D)=(4,4,16,16,6,12,30,30)$.}
\label{Ablation experiment}
\begin{tabular}{c c c c c||c c}
\hline
\makebox[0.044\textwidth]{MP} & \makebox[0.044\textwidth]{RD} & \makebox[0.044\textwidth]{$\mathbb{C}$FL1} & \makebox[0.044\textwidth]{$\mathbb{C}$FL2} & \makebox[0.044\textwidth]{$\mathbb{C}$FL3} & EE & SR \\
 \hline
$\checkmark$ & $\checkmark$ & $\checkmark$ & $\checkmark$ & $\checkmark$ & 18.84 & 177.59 \\
 \cline{6-7}
$\times$ & $\checkmark$ & $\checkmark$ & $\checkmark$ & $\checkmark$ & 17.36 & 172.08 \\
 \cline{6-7}
$\checkmark$ & $\times$ & $\checkmark$ & $\checkmark$ & $\checkmark$ & 8.87 & 128.30 \\
 \cline{6-7}
$\checkmark$ & $\checkmark$ & $\times$ & $\checkmark$ & $\checkmark$ & 13.26 & 170.23 \\
 \cline{6-7}
$\checkmark$ & $\checkmark$ & $\checkmark$ & $\times$ & $\checkmark$ & 14.46 & 155.80 \\
 \cline{6-7}
$\checkmark$ & $\checkmark$ & $\checkmark$ & $\checkmark$ & $\times$ & 14.57 & 144.74 \\
 \cline{6-7}
$\checkmark$ & $\checkmark$ & $\times$ & $\times$ & $\times$ & 8.79 & 132.61 \\
 \hline
\end{tabular}
\begin{tablenotes}
        \footnotesize
        \item MP/RD/$\mathbb{C}$FL1/$\mathbb{C}$FL2/$\mathbb{C}$FL3: message passing/residual/$\mathbb{C}$FL in Stage 1/$\mathbb{C}$FL in Stage 2/$\mathbb{C}$FL in Stage 3.
\end{tablenotes}
\end{table}

Table~\ref{Ablation experiment} presents the ablation study on the main architectural components of the proposed GNN. Removing message passing leads to only a moderate performance loss, suggesting that the reconstructed effective channels already provide informative structural priors, while message passing further enriches the extraction of graph-topological information. In contrast, removing the residual connection causes a much more pronounced degradation, since deep graph layers are more vulnerable to over-smoothing without residual fusion. When the $\mathbb{C}$FL in each stage is removed individually, the performance degrades in every case, and the degradation becomes even more significant when they are removed simultaneously. This confirms that the complex embedding decoding process is crucial in every stage of the joint optimization.

\subsection{Impact of Number of PAs}


\begin{figure}[t]
    \includegraphics[width=0.50\textwidth]{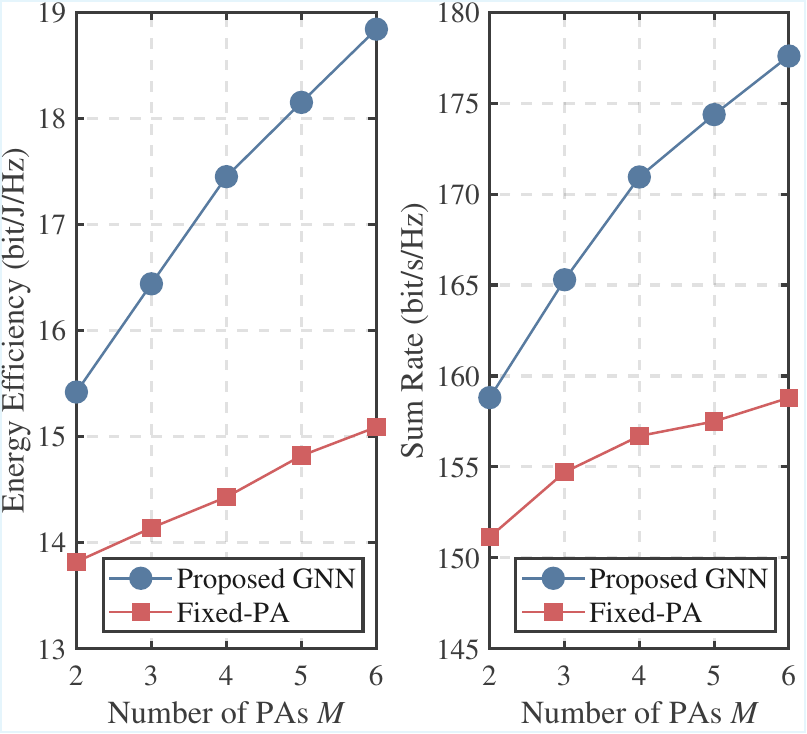}
    \caption{Impact of number of PAs on EE and SR, with $(B,R,N,L,K,S,D)=(4,4,16,16,12,30,30)$.}
    \label{antenna_impact}
\end{figure}

Figure~\ref{antenna_impact} shows that increasing the number of PAs improves both EE and SR for all compared schemes. The gain is largest when moving from a small number of PAs to a moderate one, and then gradually saturates, which is consistent with the diminishing marginal spatial DoF offered by additional elements. Moreover, the performance gap between the movable-PA schemes and the fixed-PA counterpart becomes more visible as $M$ increases, which indicates that a larger set of PAs allows the optimizer to exploit PA mobility more effectively.




\section{Conclusion}
For coordinated downlink transmission in a multi-BS multi-RIS-assisted PA system, we have presented an unsupervised three-stage GNN architecture, composed of ChanGNN, BeamGNN, and AssocGNN in cascade, for jointly optimizing PA positions, RIS phase shifts, transmit beamforming, and BS-UE association. By integrating heterogeneous and homogeneous graph representations, the proposed framework to effectively captures the structured interactions among BSs, RISs, and UEs. In addition, feasibility-preserving output mechanisms ensure valid PA placement, RIS phase normalization, power allocation, and BS-UE association throughout the end-to-end learning process. Extensive numerical results demonstrate the proposed framework’s superior performance over representative system and model baselines, together with favorable scalability and millisecond-level inference time. To the best of our knowledge, effective transmission design for PA systems under the challenging multi-BS multi-RIS setting considered in this work has not yet been reported in the open literature. Therefore, the proposed GNN provides a practically viable and pioneering solution for future wireless communications.


\begin{thebibliography}{10}

\bibitem{wang2023road}
C.-X. Wang \emph{et al.}, ``On the road to 6G: Visions, requirements, key technologies, and testbeds,'' \emph{IEEE Commun. Surv. Tutorials}, vol. 25, no. 2, pp. 905--974, Second Quart., 2023.

\bibitem{di2020smart1}
M. Di Renzo, A. Zappone, M. Debbah, M.-S. Alouini, C. Yuen, J. de Rosny, and S. Tretyakov, ``Smart radio environments empowered by reconfigurable intelligent surfaces: How it works, state of research, and the road ahead,'' \emph{IEEE J. Sel. Areas Commun.}, vol. 38, no. 11, pp. 2450--2525, Nov. 2020.

\bibitem{basar2019wireless}
E. Basar, M. Di Renzo, J. de Rosny, M. Debbah, M.-S. Alouini, and R. Zhang, ``Wireless communications through reconfigurable intelligent surfaces,'' \emph{IEEE Access}, vol. 7, pp. 116753--116773, 2019.

\bibitem{wu2021intelligent}
Q. Wu, S. Zhang, B. Zheng, C. You, and R. Zhang, ``Intelligent reflecting surface-aided wireless communications: A tutorial,'' \emph{IEEE Trans. Commun.}, vol. 69, no. 5, pp. 3313--3351, May 2021.

\bibitem{liu2021reconfigurable}
Y. Liu, X. Liu, X. Mu, T. Hou, J. Xu, M. Di Renzo, and N. Al-Dhahir, ``Reconfigurable intelligent surfaces: Principles and opportunities,'' \emph{IEEE Commun. Surv. Tutorials}, vol. 23, no. 3, pp. 1546--1577, Third Quart., 2021.

\bibitem{huang2019reconfigurable}
C. Huang, A. Zappone, G. C. Alexandropoulos, M. Debbah, and C. Yuen, ``Reconfigurable intelligent surfaces for energy efficiency in wireless communication,'' \emph{IEEE Trans. Wireless Commun.}, vol. 18, no. 8, pp. 4157--4170, Aug. 2019.

\bibitem{guan2020intelligent}
X. Guan, Q. Wu, and R. Zhang, ``Intelligent reflecting surface assisted secrecy communication: Is artificial noise helpful or not?'' \emph{IEEE Wireless Commun. Lett.}, vol. 9, no. 6, pp. 778--782, Jun. 2020.

\bibitem{yu2020robust}
X. Yu, D. Xu, and R. Schober, ``Robust and secure wireless communications via intelligent reflecting surfaces,'' \emph{IEEE J. Sel. Areas Commun.}, vol. 38, no. 11, pp. 2637--2652, Nov. 2020.

\bibitem{pan2021intelligent}
C. Pan \emph{et al.}, ``Multicell MIMO communications relying on intelligent reflecting surfaces,'' \emph{IEEE Trans. Wireless Commun.}, vol. 19, no. 8, pp. 5218--5233, Aug. 2020.

\bibitem{zhu2024movable}
L. Zhu, W. Ma, and R. Zhang, ``Movable antennas for wireless communication: Opportunities and challenges,'' \emph{IEEE Commun. Mag.}, vol. 62, no. 6, pp. 114--120, Jun. 2024.

\bibitem{new2025fluid}
W. K. New \emph{et al.}, ``A tutorial on fluid antenna system for 6G networks: Encompassing communication theory, optimization methods and hardware designs,'' \emph{IEEE Commun. Surv. Tutorials}, vol. 27, no. 4, pp. 2325--2377, Fourth Quart., 2025.

\bibitem{liu2025passtutorial}
Y. Liu, H. Jiang, X. Xu, Z. Wang, J. Guo, C. Ouyang, X. Mu, Z. Ding, A. Nallanathan, G. K. Karagiannidis, and R. Schober, ``Pinching-antenna systems (PASS): A tutorial,'' \emph{IEEE Trans. Commun.}, vol. 74, pp. 4881-4918, 2026.

\bibitem{ding2025pinching}
Z. Ding, R. Schober, and H. V. Poor, ``Flexible-antenna systems: A pinching-antenna perspective,'' \emph{IEEE Trans. Commun.}, vol. 73, no. 10, pp. 9236--9253, Oct. 2025.

\bibitem{zhu2024modeling}
L. Zhu, W. Ma, and R. Zhang, ``Modeling and performance analysis for movable antenna enabled wireless communications,'' \emph{IEEE Trans. Wireless Commun.}, vol. 23, no. 6, pp. 6234--6250, Jun. 2024.

\bibitem{zhu2024multiuser}
L. Zhu, W. Ma, B. Ning, and R. Zhang, ``Movable-antenna enhanced multiuser communication via antenna position optimization,'' \emph{IEEE Trans. Wireless Commun.}, vol. 23, no. 7, pp. 7214--7229, Jul. 2024.

\bibitem{ma2024mimo}
W. Ma, L. Zhu, and R. Zhang, ``MIMO capacity characterization for movable antenna systems,'' \emph{IEEE Trans. Wireless Commun.}, vol. 23, no. 4, pp. 3392--3407, Apr. 2024.

\bibitem{mei2024graphma}
W. Mei, X. Wei, B. Ning, Z. Chen, and R. Zhang, ``Movable-antenna position optimization: A graph-based approach,'' \emph{IEEE Wireless Commun. Lett.}, vol. 13, no. 7, pp. 1853--1857, Jul. 2024.

\bibitem{wei2025friendsfoes}
X. Wei, W. Mei, Q. Wu, Q. Jia, B. Ning, Z. Chen, and J. Fang, ``Movable antennas meet intelligent reflecting surface: Friends or foes?'' \emph{IEEE Trans. Commun.}, vol. 73, no. 11, pp. 12756--12770, Nov. 2025.

\bibitem{bjornson2014optimal}
E. Bj\"ornson, M. Bengtsson, and B. Ottersten, ``Optimal multiuser transmit beamforming: A difficult problem with a simple solution structure,'' \emph{IEEE Signal Process. Mag.}, vol. 31, no. 4, pp. 142--148, Jul. 2014.

\bibitem{cvx-iterations}
Y. Shi, L. Lian, Y. Shi, Z. Wang, Y. Zhou, L. Fu, L. Bai, J. Zhang, and W. Zhang, ``Machine learning for large-scale optimization in 6G wireless networks,'' \emph{IEEE Commun. Surv. Tutorials}, vol. 25, no. 4, pp. 2088--2132, Fourth Quart., 2023.

\bibitem{zhang2019deep}
C. Zhang, P. Patras, and H. Haddadi, ``Deep learning in mobile and wireless networking: A survey,'' \emph{IEEE Commun. Surv. Tutorials}, vol. 21, no. 3, pp. 2224--2287, Third Quart., 2019.

\bibitem{lin2019bfnn}
T. Lin and Y. Zhu, ``Beamforming design for large-scale antenna arrays using deep learning,'' \emph{IEEE Wireless Commun. Lett.}, 2019, doi: 10.1109/LWC.2019.2943466.

\bibitem{deeptx2022}
J. M. J. Huttunen, D. Korpi, and M. Honkala, ``DeepTx: Deep learning beamforming with channel prediction,'' \emph{IEEE Trans. Wireless Commun.}, vol. 22, no. 3, pp. 1855-1867, Mar. 2023.

\bibitem{kang2024maDL}
J.-M. Kang, ``Deep learning enabled multicast beamforming with movable antenna array,'' \emph{IEEE Wireless Commun. Lett.}, vol. 13, no. 7, pp. 1848--1852, Jul. 2024.

\bibitem{shen2020graph}
Y. Shen, Y. Shi, J. Zhang, and K. B. Letaief, ``Graph neural networks for scalable radio resource management: Architecture design and theoretical analysis,'' \emph{IEEE J. Sel. Areas Commun.}, vol. 39, no. 1, pp. 101--115, Jan. 2021.

\bibitem{shen2023gnn}
Y. Shen, J. Zhang, S. H. Song, and K. B. Letaief, ``Graph neural networks for wireless communications: From theory to practice,'' \emph{IEEE Trans. Wireless Commun.}, vol. 22, no. 5, pp. 3554--3569, May 2023.

\bibitem{li2024gnn}
Y. Li, Y. Lu, B. Ai, O. A. Dobre, Z. Ding, and D. Niyato, ``GNN-based beamforming for sum-rate maximization in MU-MISO networks,'' \emph{IEEE Trans. Wireless Commun.}, vol. 23, no. 8, pp. 9251--9264, Aug. 2024.

\bibitem{wang2024gnn}
Q. Wang, Y. Lu, W. Chen, B. Ai, Z. Zhong, and D. Niyato, ``GNN-enabled optimization of placement and transmission design for UAV communications,'' \emph{IEEE Trans. Veh. Technol.}, vol. 73, no. 11, pp. 16789--16802, Nov. 2024.

\bibitem{he2025fasnet}
C. He, Y. Lu, W. Chen, B. Ai, K.-K. Wong, and D. Niyato, ``Graph neural network enabled fluid antenna systems: A two-stage approach,'' \emph{IEEE Trans. Veh. Technol.}, vol. 74, no. 10, pp. 16625--16629, Oct. 2025.

\bibitem{xie2025graph}
X. Xie, Y. Lu, and Z. Ding, ``Graph neural network enabled pinching antennas,'' \emph{IEEE Wireless Commun. Lett.}, 2025, doi: 10.1109/LWC.2025.3584919.

\bibitem{gpass2025}
J. Guo, Y. Liu, and A. Nallanathan, ``GPASS: Deep learning for beamforming in pinching-antenna systems (PASS),'' arXiv preprint arXiv:2502.01438, 2025.

\bibitem{rispassgnn2025}
C. He, Y. Lu, Y. Xu, C.-Y. Chi, B. Ai, and A. Nallanathan, ``RIS-assisted downlink pinching-antenna systems: GNN-enabled optimization approaches,'' arXiv preprint arXiv:2511.20305, 2025.

\bibitem{risassoc2021}
S. Liu, P. Ni, R. Liu, Y. Liu, M. Li, and Q. Liu, ``BS-RIS-user association and beamforming designs for RIS-aided cellular networks,'' in \emph{Proc. IEEE/CIC Int. Conf. Commun. China (ICCC)}, Xiamen, China, 2021, pp. 563-568.

\bibitem{bereyhi2025downlink}
A. Bereyhi, S. Asaad, C. Ouyang, Z. Ding, and H. V. Poor, ``Downlink beamforming with pinching-antenna assisted MIMO systems,'' in \emph{Proc. IEEE Int. Conf. Commun. Workshops (ICC Workshops)}, Montreal, QC, Canada, 2025, pp. 1-6.

\bibitem{xu2025jointpass}
X. Xu, X. Mu, Y. Liu, and A. Nallanathan, ``Joint transmit and pinching beamforming for pinching antenna system (PASS): Optimization-based or learning-based?,'' \emph{IEEE Trans. Wireless Commun.}, vol. 25, pp. 11449-11464, 2026.

\bibitem{le2024starrisgnn}
H. A. Le, T. Van Chien, and W. Choi, ``Graph neural network-based active and passive beamforming for distributed STAR-RIS-assisted multi-user MISO systems,'' \emph{IEEE Trans. Commun.}, vol. 73, no. 10, pp. 9299-9312, Oct. 2025.

\bibitem{liu2025multirisassoc}
M. Liu, C. Huang, A. Alhammadi, M. Di Renzo, M. Debbah, and C. Yuen, ``Beamforming design and association scheme for multi-RIS multi-user mmWave systems through graph neural networks,'' \emph{IEEE Trans. Wireless Commun.}, vol. 24, no. 9, pp. 7940-7954, Sept. 2025.


\bibitem{hyb}
 W. Guo et al., ``Hybrid MRT and ZF learning for energy-efficient transmission in multi-RIS-assisted networks," \emph{IEEE Trans. Veh. Technol.}, vol. 73, no. 8, pp. 12247-12251, Aug. 2024.

\bibitem{Gumbel-Softmax}
E. Jang, S. Gu, and B. Poole, ``Categorical reparameterization with Gumbel-Softmax,'' in \emph{Proc. ICLR}, 2017, pp. 1920-1931.

\bibitem{han}
X. Wang et al., `Heterogeneous graph attention network,” in \emph{Proc. ACM WWW}, pp. 2022-2032, 2019.

\bibitem{complex NN}
C. Trabelsi, O. Bilaniuk, Y. Zhang, D. Serdyuk, S. Subramanian, J. F. Santos, S. Mehri, N. Rostamzadeh, Y. Bengio, and C. J Pal, ``Deep complex networks,'' in \emph{Proc. ICLR}, 2018, pp.1-19.

\bibitem{gat} P. Velickovic, G. Cucurull, A. Casanova, A. Romero, P. Lio, and Y. Bengio, `Graph attention networks,'' in \emph{Proc. ICLR}, pp. 1–12, 2018.

\bibitem{kaiming_normal}
K. He, X. Zhang, S. Ren, and J. Sun, ``Delving deep into rectifiers: Surpassing human-level performance on ImageNet classification," in \emph{Proc. ICCV}, pp. 1026-1034, 2015.
 
\bibitem{Adam}
D. P. Kingma and J. Ba, ``Adam: A method for stochastic optimization,” in \emph{Proc. ICLR}, pp. 1–15, Feb. 2015.

\end{thebibliography}
\end{document}